\documentclass[preprint,showpacs,preprintnumbers,amsmath,amssymb,showpacs,11pt]{revtex4}

\usepackage{graphicx}
\usepackage{dcolumn}
\usepackage{bm}
\usepackage{amssymb}

\begin{document}

\title{Discovering Relic Gravitational Waves in Cosmic Microwave Background
Radiation\footnote{Based on the invited lecture at the 1-st J. A. Wheeler 
School on Astrophysical Relativity, June 2006, Italy.}}

\vspace{3.5cm}

\author{L. P. Grishchuk\footnote{e-mail address: grishchuk@astro.cf.ac.uk}}

\affiliation{School~of~Physics~and~Astronomy,~Cardiff~University,~Cardiff~
CF243AA,~UK}
\affiliation{Sternberg Astronomical Institute, Moscow State University, 
Moscow 119899,~Russia}

\date{\today}


\begin{abstract}
The authority of J. A. Wheeler in many areas of gravitational physics is 
immense, and there is a connection with the study of relic gravitational 
waves as well. I begin with a brief description of Wheeler's influence on
this study. One part of the paper is essentially a detailed justification 
of the very existence of relic gravitational waves, account of their 
properties related to the quantum-mechanical origin, derivation of the 
expected magnitude of their effects, and reasoning why they should be 
detectable in the relatively near future. This line of argument includes 
the comparison of relic gravitational waves with density perturbations of 
quantum-mechanical origin, and the severe criticism of methods and predictions 
of inflationary theory. Another part of the paper is devoted to active 
searches for relic gravitational waves in cosmic microwave background 
radiation (CMB). Here, the emphasis is on the temperature-polarization $TE$ 
cross-correlation function of CMB. The expected numerical level of the 
correlation, its sign, statistics, and the most appropriate interval of 
angular scales are identified. Other correlation functions are also considered. 
The overall conclusion is such that the observational discovery of relic 
gravitational waves looks like the matter of a few coming years, rather than 
a few decades.     
\end{abstract}



\maketitle



\section{\label{sec:introduction}Introduction}

It is my honor and pleasure to be a lecturer at the first course of 
J. A. Wheeler School on Astrophysical Relativity. Wheeler is one of the 
founding fathers 
of the field of gravitational physics, and many of us are strongly 
influenced by his work and his personality. In particular, we often find 
inspiration in the great textbook of him and his colleagues \cite{MTW}. 

I was fortunate to speak with J. A. Wheeler right before the publication 
of my first papers on relic gravitational waves \cite{gr74}. That 
conversation helped me to shape my views on the subject. Being a young 
researcher, I was somewhat nervous about evaluation of my work by the 
towering scientist, but to my relief, Wheeler quickly understood the work 
and agreed with it. In what follows, we will be discussing relic gravitational 
waves systematically and in details, but I would like to start from describing 
my initial doubts and worries, and how Wheeler helped me to see them in a
different light.

I showed that the wave-equation for a gravitational wave $h(\eta, {\bf x}) =
[\mu(\eta)/a(\eta)] e^{i {\bf n}{\cdot {\bf x}}}$ in a homogeneous isotropic
universe with the scale factor $a(\eta)$ can be written as a 
Schrodinger-like equation
\begin{eqnarray}
{\mu}^{''} + {\mu}\left[n^{2} - \frac{a''}{a}\right] =0.
\label{meq}
\end{eqnarray}
It follows from this equation that the fate of the wave with the
wavenumber $n$ depends on the comparative values of 
$n^2$ and the effective potential $U(\eta)= a^{''}/a$. If $n^2$ is much 
larger than $|U(\eta)|$, the wave does not feel the potential 
and propagates with the adiabatically changing amplitude $h \propto 1/a$. 
In the opposite limit, the interaction with the potential is strong and the 
wave changes dramatically. The amplitude of the initial wave gets amplified 
over and above the adiabatic law $h \propto 1/a$, and at the same time a 
wave propagating in the opposite direction is being created. This process 
results effectively in the production of standing waves. I called this 
phenomenon the superadiabatic amplification. Over the years, other 
cosmological wave equations were also modeled on Eq.(\ref{meq}). In terms
of the variable $h(\eta)$, where $h(\eta) = \mu(\eta)/a(\eta)$, Eq.(\ref{meq})
has the form
\begin{equation}
\label{heq}
h^{\prime \prime} +2\frac{a^{\prime}}{a} h^{\prime}+n^2 h=0.
\end{equation}

While analyzing various scale factors $a(\eta)$ and potentials $U(\eta)$, I
was imagining them as being `drawn by a hand'. My first example was the 
potential $U(\eta)$ architypical for quantum mechanics - a rectangular 
barrier. This means that $U(\eta) = const$ in some interval of $\eta$-time 
between $\eta_a$ and $\eta_b$, while $U(\eta)=0$ outside this 
interval. I have shown that the waves interacting with this barrier are 
necessarily amplified. However, the postulated $U(\eta)$ caused some 
concerns.

It is easy to make $U(\eta) =0$ before $\eta_a$ and after $\eta_b$.
Indeed, if $a^{''}/a = 0$, the scale factor 
$a(\eta)$ is either a constant, like in a flat Minkowski world, or is 
proportional to $\eta$, like in a radiation-dominated universe. It is not 
difficult to imagine that the Universe was radiation-dominated before and 
after some crucial interval of evolution. However, if it is assumed
that $a^{''}/a = const$ between $\eta_a$ and $\eta_b$, it is not 
easy to find a justification for this evolution, as the scale factor 
$a(\eta)$ should depend exponentially on $\eta$-time in this interval. 
In terms of $t$-time, which is related to $\eta$-time by $c~dt= a~d\eta$,
the scale factor $a(t)$ should be proportional to $t$. The Einstein 
equations allow this law of expansion, but they demand that the      
`matter' driving this evolution should have the effective equation of 
state $p= - (1/3) \epsilon$. Some other potentials $U(\eta)$ 
drawn by a hand do also require strange equations of state. 

Having shown the inevitability of superadiabatic amplification, 
as soon as $a^{''}/a \neq 0$, I was somewhat embarassed by the fact that 
in some parts of my study I operated with scale factors 
drawn by a hand and driven by matter with unusual equations of state. 
Although equations of state with negative pressure had already been 
an element of cosmological research, notably in the work of 
A. D. Sakharov \cite{sakh65}, I feared that Wheeler may dislike this
idea and may say `forget it'. To my surprise, he accepted
the approach and even suggested a wonderful name: the `engine-driven
cosmology'. The implication was that although we may not know the nature 
of the `engine' which drives a particular $a(\eta)$, this knowledge is not, 
for now, our high priority. Being inspired by Wheeler's attitude, I hurried 
to include the notion of the engine-driven cosmology, with reference to 
Wheeler, in the very first paper on the subject \cite{gr74}. 

It is interesting to note that E. Schrodinger \cite{schr} felt uneasy about
wave solutions in an expanding universe. (I became aware of his work much 
after the time of my first publications.) Schrodinger identified the crucial
notion of the ``mutual adulteration of positive and negative frequency 
terms in the course of time". He was thinking about electromagnetic waves 
and he called the prospect of photon creation in an expanding universe 
an ``alarming phenomenon". From the position of our present knowledge, 
we can say that Schrodinger was right to be doubtful. Indeed, electromagnetic 
waves cannot be amplified and photons cannot be created in a nonstationary 
universe. Even though the wavelengths
of electromagnetic and gravitational waves change in exactly the same
manner, their interactions with external gravitational field are drastically
different. The corresponding effective potential $U(\eta)$ in the Maxwell 
equations is strictly zero and the ``alarming phenomenon" does not take
place. All physical fields are tremendously `stretched' by expansion, but 
only some of them are being amplified. 

In his paper, Schrodinger was operating with a variant of scalar 
electrodynamics, wherein the coupling of scalar fields to gravity is
ambiguous and can be chosen in such a way that the wave equation becomes
identical to Eq.(\ref{meq}), making the amplification of scalar
waves possible (for more details, see introductory part to 
Ref.\cite{gr93prd}). L. Parker \cite{parker} undertook a systematic study
of the quantized version of test scalar fields in FLRW (Friedmann-
Lemaitre-Robertson-Walker) cosmologies. As for the gravitational waves,
there is no ambiguity in their coupling to gravity since the coupling 
follows directly from the Einstein equations. As we see, the 
``alarming phenomenon" does indeed take place for gravitational 
waves \cite{gr74}. (The authors of publications preceeding Ref.\cite{gr74} 
explicitely denied the possibility of graviton creation in FLRW universes. 
In the end, it was only Ya. B. Zeldovich who wrote to me: ``Thank you for 
your goal in my net".)

It is important to realize that the possibility of generation of relic 
gravitational waves relies only on the validity of general relativity and 
quantum mechanics. The governing principles are part of the well-understood 
and tested physics. The underlying equation (\ref{meq}) admits an analogy
with the Schrodinger equation (outlined above), but it can also be viewed 
as an equation for a classical oscillator with variable frequency. 
The phenomenon of superadiabatic (parametric) amplification of the waves' 
zero-point quantum oscillations is at the heart of the cosmological 
generating mechanism. In order to better appreciate this phenomenon, 
we shall briefly review the closely related laboratory-type problem of 
parametric amplification in a classical pendulum.

\begin{figure}
\begin{center}
\includegraphics[width=9cm]{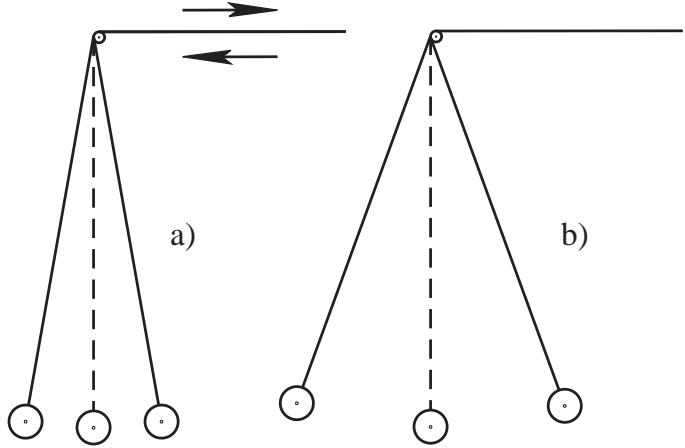}
\end{center}
\caption{Parametric amplification. (a) Variation of the length of the pendulum.
(b) Increased amplitude of oscillations.}
\label{LPG:fig1}
\end{figure}

Let us consider an ideal pendulum hanging in a constant gravitational field 
characterized by the free-fall acceleration $g$ (see Fig.\ref{LPG:fig1}).
The frequency of the oscillator is given by $\omega_0 = \sqrt{g/l}$, 
where $l$ is the length of the pendulum,
\[
\ddot{x} + \frac{g}{l} x =0.
\]
The amplitude of oscillations can be 
enhanced either by force acting directly on the pendulum's mass or by an
influence which changes a parameter of the oscillator -- in this case, its
frequency. The equation for small horizontal 
displacements $x(t)$ of the oscillator's mass takes the form 
\begin{equation}
\ddot{x} + \omega^{2}(t) x =0.
\label{pareq}
\end{equation}

The simplest parametric intervention 
makes the length $l$ time-dependent, as shown in Fig.\ref{LPG:fig1}a.
Then, $\omega^2(t)$ in Eq.(\ref{pareq}) is 
$\omega^2(t) = (g-\ddot{l})/l(t)$. Note that even if the gravitational 
accelaration $g$ remains constant, it still gets modified by the 
acceleration term $\ddot{l}$ arising due to the variation of 
length $l(t)$. So, the correct $\omega^2(t)$ differs from the naive 
expectation $\omega^2(t) = g/l(t)$. (For more details about parametrically 
excited oscillators, see \cite{byz}.) In general, 
both $g(t)$ and $l(t)$ are functions of time, in which case
$\omega^2(t) = [g(t)-\ddot{l}]/l(t)$ and Eq.(\ref{pareq}) reads
\begin{equation}
\ddot{x} + x \left[\frac{g(t)}{l(t)} - \frac{\ddot{l}}{l} \right] =0.
\label{pareq2}
\end{equation}

If the external influence on the oscillator is very slow, i. e. 
$\omega_0 \gg |{\dot{\omega}}(t)/\omega(t)|$, 
the ratio of the slowly changing energy $E(t)$ and frequency $\omega(t)$,
\begin{equation}
\label{adinv}
\frac{E}{\hbar \omega} = N,
\end{equation} 
will remain constant. This ratio is called an adiabatic invariant 
\cite{LLmech}. The quantity $N$ is also a `number of quanta' in a 
classical oscillator (see paper by Ya. B. Zeldovich, signed by a pseudonym, 
on how quantum mechanics helps understand classical mechanics \cite{zeld}).

On the other hand, if the oscillator was subject to some interval of 
appropriate parametric influence, the amplitude of oscillations and the 
number of quanta $N$ will significantly increase, as shown in 
Fig.\ref{LPG:fig1}b. To get a significant effect, the function 
$\omega(t)$ does not have to be periodic, but in its Fourier spectrum 
there should be enough power at frequencies around $\omega_0$. The final 
frequency does not need to differ from the initial $\omega_0$.

One can notice the striking analogy between Eq.(\ref{meq}) and 
Eq.(\ref{pareq2}). This analogy extends further if one goes over from 
the dispacement $x(t)$ to the dimensionless angle variable $\phi(t)$ 
realated to $x(t)$ by $\phi(t) = x(t)/l(t)$. (Compare with the 
cosmological relationship $h(\eta) = \mu(\eta)/a(\eta)$.) 
Then, Eq.(\ref{pareq2}) takes the form
\begin{equation}
\label{pareq3}
\ddot{\phi} +2\frac{\dot{l}}{l} \dot{\phi} + \frac{g(t)}{l(t)} \phi=0,
\end{equation}
which is strikingly similar to Eq.(\ref{heq}). In cosmological equations, 
the analog of the ratio $g(t)/l(t)$ is $n^2$.

There are two lessons to be learned from this discussion. First, the 
necessary condition for a significant amplification of the wave is 
the availability of a regime where the characteristic time $|a/a'|$ of 
variation of the external gravitational field (represented by the scale 
factor $a(\eta)$) becomes comparable and much shorter than the wave 
period $2 \pi/n$,
\begin{equation}
n \ll \left|\frac{a'(\eta)}{a(\eta)}\right|.
\label{cond}
\end{equation}
Before and after this regime, the wave may be a high-frequency wave,
that is, it may satisfy the condition 
$n \gg \left|{a'(\eta)}/{a(\eta)}\right|$. But if during some 
interval of evolution the opposite condition (\ref{cond}) is 
satisfied, and $a^{''}/a \neq 0$, the wave will be superadiabatically 
amplified, regardless of whether the model universe is expanding or 
contracting \cite{gr74}.

Second, a classical pendulum should 
initially be in a state of oscillations -- excited state -- in order 
to have a chance to be amplified by an external `pump' influence. Otherwise, 
if it is initially at rest, i.e. hanging stright down, the time-dependent 
change of $l$ or $g$ will not excite the oscillator, and the energy of 
oscillations will remain zero. However, in the quantum world, even if the 
oscillator is in the state of lowest energy (ground, or vacuum, state) it 
will inevitably possess the zero-point quantum oscillations. One can think of 
these zero-point quantum oscillations as those that are being amplified 
by the external pump (assuming, of course, that the oscillator is coupled 
to the pump). This is where the quantum mechanics enters the picture. The
initial ground state of the parametrically excited oscillator evolves into 
a multi-quantum state, and the mean number of quanta grows, when 
condition (\ref{cond}) is satisfied \cite{gr93}. There is no quantum state 
lower than the ground state, so the engine-driven cosmology will necessarily 
bring the properly coupled quantum oscillator into an excited state.  

The concept of the engine-driven cosmology remains perfectly adequate for
the present-day research. We do not know what governed the scale factor 
of the very early Universe. It could be a lucky version of the scalar field 
(inflation) or something 
much more sophisticated and dictated by the `theory of everything'. 
The lack of this knowledge is not important for the time being. By observing 
relic gravitational waves we may not be able to determine at once the nature of 
the cosmological `engine', but we will be able to determine the behaviour of
the early Universe's scale factor. Therefore, we will gain unique information 
about the `birth' of the Universe and its very early dynamical evolution.

\section{\label{sec:cosos}Cosmological Oscillators}

The physics of laboratory-type oscillators have direct relevance to cosmological
oscillators. In cosmology, we normally consider a universe filled with some 
matter and slightly perturbed in all constituents. It is convenient to write 
the perturbed metric of a flat FLRW universe in the form
\begin{eqnarray}
ds^{2} = -c^{2} dt^{2} + a^{2}(t)(\delta_{ij} + h_{ij}) dx^idx^j=
a^{2}(\eta)\left[ - d\eta^{2} + (\delta_{ij} + h_{ij}) dx^idx^j\right].
\label{FRWmetric}
\end{eqnarray}
The six functions $h_{ij}\left(\eta,{\bf x}\right)$ can be expanded over
spatial Fourier harmonics $e^{\pm i {\bf n\cdot x}}$, where ${\bf n}$ is a 
dimensionless time-independent wave-vector,
\begin{eqnarray}
h_{ij}\left(\eta,{\bf x}\right) = \frac{\mathcal{
C}}{(2\pi)^{3/2}}\int\limits_{-\infty}^{+\infty}~d^{3}{\bf
n}\frac{1}{\sqrt{2n}} \sum_{s=1,2} \left[\stackrel{s}{p}_{ij}({\bf
n}) \stackrel{s}{h}_n\left(\eta\right)e^{i{\bf n\cdot x}}
\stackrel{s}{c}_{\bf n}+ {\stackrel{s}{p}_{ij}}^{*}({\bf n})
{\stackrel{s}{h}_n}^{*}(\eta)e^{-i{\bf n\cdot x}}
\stackrel{s}{c}_{\bf n}^{*}\right].
\label{fourierh}
\end{eqnarray}

This representation requires some explanations. The mode functions
$\stackrel{s}{h}_n\left(\eta\right)$ obey differential equations 
that follow from the perturbed Einstein equations (as an example,
look at Eq.(\ref{meq})). The wavelength
of the mode ${\bf n}$ is given by $\lambda = 2\pi a /n$, where the 
wavenumber $n$ is $n=(\delta_{ij}n^in^j)^{1/2}$. It is convenient to 
take today's scale factor $a(\eta_R)$ to be equal to the `size of the 
Universe', that is, $a(\eta_R) = 2l_H$, where $l_H = c/ H_0$ is today's 
Hubble radius and $H_0=H(\eta_R)$ is today's Hubble parameter. Then, 
for a fixed moment of time, today, we can also write the laboratory-type
expression for the same wavelenghth $\lambda$: $\lambda= 2 \pi/k$, 
where $k$ has the dimensionality of inverse length and is related to the
dimensionless $n$ by $k = n/2 l_H$. The wave whose length is 
equal to today's `size of the Universe' has $n=2\pi$. It is assumed that
wavelengths can always be measured by unchangeable laboratory standards. 

By systematically writing ${\bf n}$ and distinguishing it from ${\bf k}$ 
we essentially follow the original motivations of 
E. M. Lifshitz (see, for example, \cite{LL}). In contrast, 
some modern authors think that Lifshitz needs to be `simplified' and 
`modernized'. They use only one letter ${\bf k}$ and write 
${\bf k}$ everywhere where Lifshitz was writing ${\bf n}$. This created 
quite a mess in wavenumbers and wavelengths. Among them you will see 
physical, non-physical, coordinate, comoving, proper, etc.

Moving to complex Fourier coefficients $\stackrel{s}{c}_{\bf n}$, 
$\stackrel{s}{c}_{\bf n}^{*}$ we note that they are some particular numbers, 
if the left-hand-side (l.h.s) of Eq.(\ref{fourierh}) is a deterministic, 
even if arbitrarily complicated, function. In the rigorous quantum-mechanical  
version of the theory, these coefficients will be promoted to the status of 
quantum-mechanical annihilation and creation operators 
$\stackrel{s}{c}_{\bf n}$, $\stackrel{s}{c}_{\bf n}^{\dag}$ acting on
some quantum states. In the CMB applications, we will treat, for simplicity,
$\stackrel{s}{c}_{\bf n}$, $\stackrel{s}{c}_{\bf n}^{*}$ as random 
numbers taken from some probability distributions. The factor $1/\sqrt{2n}$
in Eq.(\ref{fourierh}) is a useful insertion inspired by quantum field 
theories. The normalization constant $\cal C$ will be discussed later.

The gravitational field polarization tensors $\stackrel{s}{p}_{ij}({\bf n})$
deserve special attention. As we shall see below, the polarization 
properties of the CMB radiation -- our final destination -- are intimately
connected with the structure of these metric polarization tensors. Polarization
of CMB and polarization of metric perturbations  
is not simply a coincidence in the usage of the word polarization.
The polarization tensors $\stackrel{s}{p}_{ij}({\bf n})$
have different forms depending on whether the functions
$h_{ij}(\eta, {\bf x})$ represent gravitational waves, rotational
perturbations, or density perturbations. Each class of these perturbations
have two polarization states, so $s= 1,2$ for each of them. In what follows 
we will be considering gravitational waves and density perturbations.   

In the case of gravitational waves, two independent linear
polarization states can be described by two real polarization
tensors
\begin{eqnarray}
\stackrel{1}{p}_{ij}({\bf n})=l_il_j - m_im_j,~~
\stackrel{2}{p}_{ij}({\bf n})=l_im_j+m_il_j,
\label{ptensors}
\end{eqnarray}
where spatial vectors $({\bf l},{\bf m},{\bf n}/n)$ are unit and
mutually orthogonal vectors. The polarization tensors
(\ref{ptensors}) satisfy the conditions
\begin{eqnarray}
\stackrel{s}{p}_{ij}\delta^{ij}=0,~~~\stackrel{s}{p}_{ij}n^{i}=0,~~~
\stackrel{s'}{p}_{ij}\stackrel{s}{p}{}^{ij}=2\delta_{s's}.
\label{orthpol}
\end{eqnarray}
Two circular polarization states are described by
\begin{eqnarray}
\stackrel{L}{p}_{ij}=\frac{1}{\sqrt 2}\left(\stackrel{1}{p}_{ij}+
i \stackrel{2}{p}_{ij}\right), ~~~~~
\stackrel{R}{p}_{ij}=\frac{1}{\sqrt 2}\left(\stackrel{1}{p}_{ij}-
i \stackrel{2}{p}_{ij}\right).
\label{pctensors}
\end{eqnarray}
The left and right polarizations interchange under a coordinate reflection 
(altering the sign of $l^i$ or $m^i$). In other words, gravitational waves
can support a chirality, or `handedness'.

In the case of density perturbations, the polarization tensors are
\begin{eqnarray}
\stackrel{1}{p}_{ij}=\sqrt{\frac{2}{3}} \delta_{ij},~~
\stackrel{2}{p}_{ij}= -\sqrt{3} \frac{n_in_j}{n^{2}}+
\frac{1}{\sqrt{3}}\delta_{ij}.
\label{ptensors4}
\end{eqnarray}
These polarization tensors satisfy the last of the conditions
(\ref{orthpol}). The polarization tensors (\ref{ptensors4}) remain 
unchanged under coordinate mirror reflections, so density perturbations
cannot support handedness.

It is important to stress that from the position of general relativity,
cosmological density perturbations represented by metric perturbations
$h_{ij}$ with polarization structure (\ref{ptensors4}), can be viewed
as `scalar', or spin-0, gravitational waves. Indeed, although
in the absense of matter, i.e. for $T_{\mu \nu}=0$, the linearised Einstein
equations admit spin-0 wave solutions, i.e. solutions with the structure 
(\ref{ptensors4}), these solutions do not carry energy, do not affect 
the relative motion of test masses, and can be nullified by coordinate 
transformations. It is only spin-2 wave solutions, i.e. solutions with the 
structure (\ref{ptensors}), that carry energy, affect the relative 
motion of test masses, and cannot be removed by coordinate transformations. 
These spin-2 solutions are called gravitational waves. 

However, in cosmology, 
the spin-0 solutions survive and become non-trivial, 
as soon as metric perturbations with the structure (\ref{ptensors4}) 
are supported by non-vanishing matter perturbations, that is, when 
$\delta T_\mu^{\nu} \neq 0$. These solutions, which unite gravitational 
field and matter perturbations, are called cosmological density 
perturbations. Therefore, cosmological density perturbations
and cosmological gravitational waves, although separate from the point of 
view of algebraic classification of the tensor $h_{ij}$, are not entirely
disconnected entities that should be treated by different theories.
On the contrary, they should be viewed as parts of a common set of 
gravitational (metric) degrees of freedom. This can be regarded as a
physical principle that will later guide our choice of initial conditions
for density perturbations. 

One more comment is in order. In our presentation, we will be consistently 
using the class of synchronous coordinate systems (\ref{FRWmetric}), that 
is, we assume that $h_{00}=0$ and $h_{0i}=0$. We are not losing anything in 
terms of physics but we gain significantly in terms of technical 
simplicity and universality of our approach to gravitational waves and
density perturbations. In principle, one can work in arbitrary coordinates,
assuming that all components of metric perturbations, 
including $h_{00}$ and $h_{0i}$, are non-zero. This will not bring you any 
real advantages, but will complicate calculations and can mislead you in 
issues of interpretation. Especially if you attempt to compare, say, 
gravitational waves descibed in synchronous coordinates with density 
perturbations described in the `Newtonian-gauge' or other `gauge' 
coordinates. Nevertheless, since various gauges and gauge-invariant 
formalisms are popular in contemporary literature, we will 
indicate, where necessary, how our formulas would be modified  
had we used arbitrary coordinates.

\section{\label{sec:quantgw}Quantization of Gravitational Waves}

Cosmological gravitational waves exist in the absence of matter perturbations,
i.e. for $\delta T_\mu^{\nu} = 0$. For each wavenumber $n$ and polarization 
state $s=1,2$ the
mode functions $\stackrel{s}{h}_n(\eta)a(\eta)=
\stackrel{s}{\mu}_n\left(\eta\right)$ satisfy the familiar equation
(\ref{meq}). We assume that each of gravitational-wave oscillators ${\bf n}$ 
was initially, at some $\eta = \eta_0$, in its ground state. 
Although certain oscillators could be somewhat excited without violating 
our perturbative assumptions, we do not see physical justification for
such non-vacuum initial states.

There is no such thing as the ground (vacuum) state without explicitely 
indicating the Hamiltonian for which the state is the ground state. Shortly, 
we will explicitly 
write down the Hamiltonian for gravitational waves (and, later, for 
density perturbations too). Specifying the Hamiltonian will also make more 
precise the concept of normalization of the initial mode 
functions to the zero-point quantum oscillations, or in other words the 
normalization to a `half of the quantum in each mode'. 

It is important to remember, however, that, qualitatively, we 
already know the answer \cite{gr74}. The energy of a gravitational wave 
with wavelength $\lambda_0$, contained in a volume 
$(\lambda_0/2)^3$, is equal to a half of the quantum, i.e. $N = 1/2$ in 
Eq.(\ref{adinv}), if the amplitude of the wave is at the level 
\begin{equation}
h_{i}(n) \approx \sqrt{\frac{G \hbar}{c^3}} \frac{n}{a_0} \approx 
\frac{l_{Pl}}{\lambda_0},
\label{iampl}
\end{equation}    
where $l_{Pl} = \sqrt{G \hbar/c^3}$, $a_0 = a(\eta_0)$ and 
$\lambda_0 = 2\pi a_0/n$. Eq.(\ref{iampl}) defines the initial vacuum spectrum 
of the gravitational wave (g.w.) amplitudes: $h_{i}(n) \propto n$. 

Shifting the initial time $\eta_0$ up to the boundary between the
adiabatic and superadiabatic regimes at $\eta =\eta_i$, we derive the 
estimate $h_{i} \sim l_{Pl}/ \lambda_i \sim l_{Pl}H_i/c$. Then, we
use the constancy of the metric amplitude $h$ throughout the 
long-wavelength (superadiabatic) regime $n \ll a'/a$, Eq.(\ref{cond}), 
that is, the regime in which the wave is `under the barrier $a'/a$'. The 
constancy of $h$ follows from the constancy of the dominant (first) 
term in the approximate long-wavelength solution \cite{gr74} to the
equations (\ref{meq}), (\ref{heq}): 
\begin{equation}
\label{lwapp}
\frac{\mu}{a} = C_1 +C_2 \int \frac{d\eta}{a^2}.
\end{equation}
This allows us to write $h_f \approx h_i$, where $h_f$ is the estimate 
of $h$ at the end of the superadiabatic regime. After having emerged
from `under the barrier $a'/a$' the wave will again behave adiabatically.

\begin{figure}
\begin{center}
\includegraphics[width=9cm]{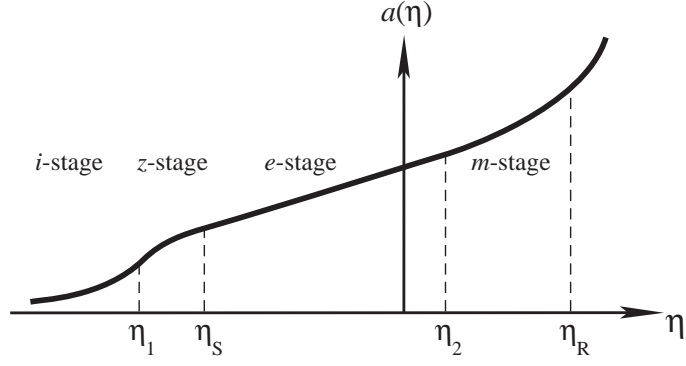}
\end{center}
\caption{A typical scale factor $a(\eta)$ as a function of time from the 
era of imposing the initial conditions, $i$-stage, and up to the present time.}
\label{scalef}
\end{figure}

\begin{figure}
\begin{center}
\includegraphics[width=9cm]{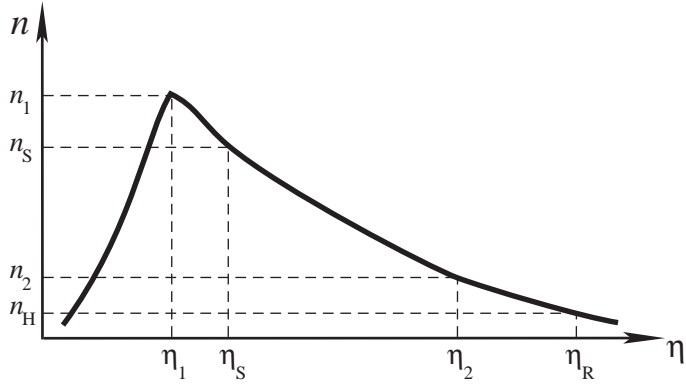}
\end{center}
\caption{The barrier $a'/a$ built from the the scale factor $a(\eta)$ 
of Fig.\ref{scalef} versus crucial wavenumbers $n$ defined by this barrier.}
\label{a'a}
\end{figure}

\begin{figure}
\begin{center}
\includegraphics[width=15cm]{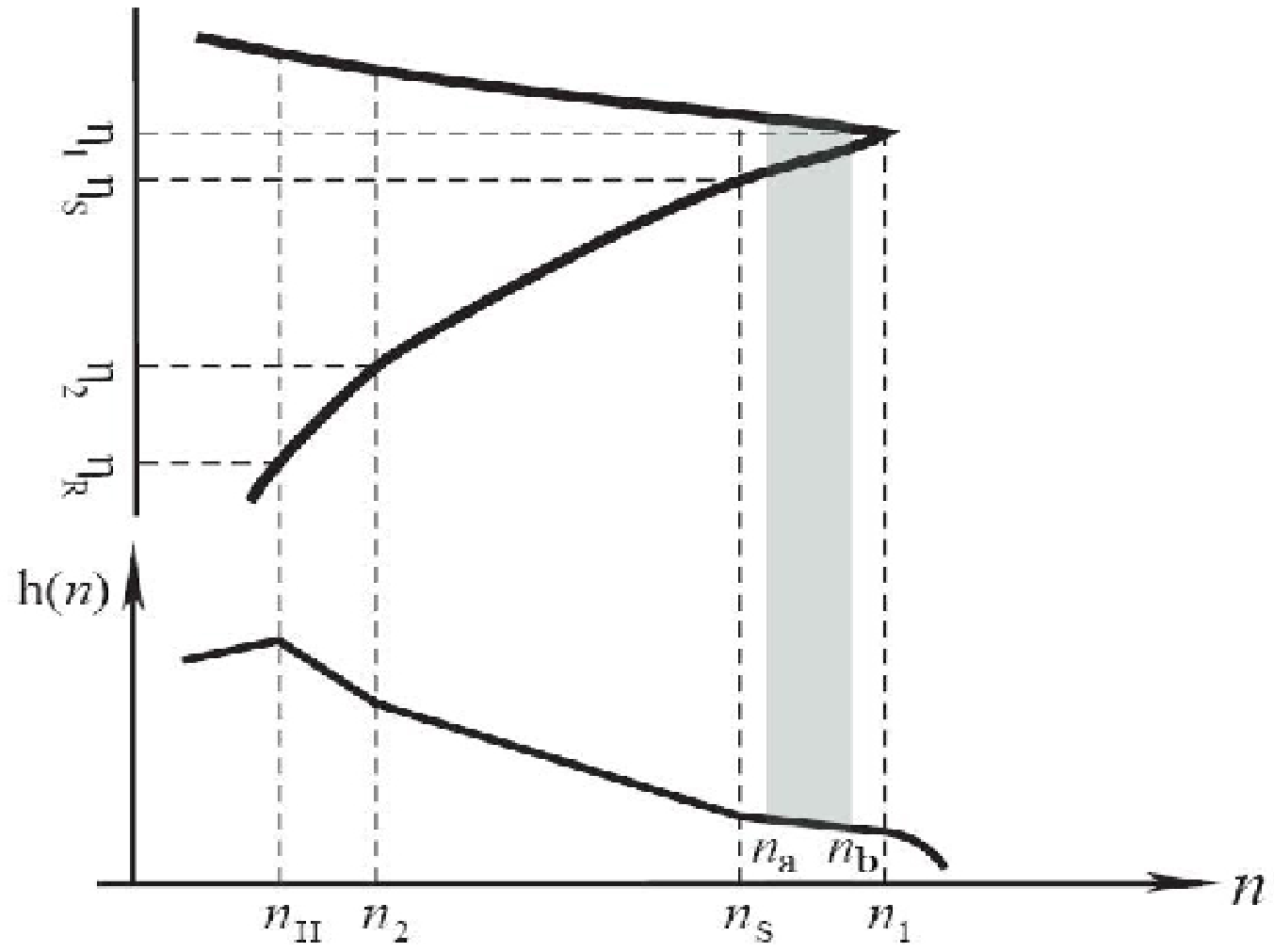}
\end{center}
\caption{The characteristic gravitational-wave amplitude $h(n)$ today as a 
function of frequency $n$. The size of the amplitude is mostly determined by 
the thickness of the barrier $a'/a$ in the place, where it is traversed by the 
wave with a given wavenumber $n$. The barrier from Fig.\ref{a'a} is shown 
at the top of the figure.}
\label{charam}
\end{figure}

Using initial conditions (\ref{iampl}) and evolving classical mode 
functions through all the barriers and intervals of adiabatic 
evolution, one can derive today's metric amplitudes as a function of 
frequency, i.e. today's amplitude spectrum. For a typical engine-driven 
expanding cosmology shown in Fig.\ref{scalef}, and for its associated 
barrier $a'/a$ shown in 
Fig.\ref{a'a}, one expects to arrive at today's spectrum qualitatively 
shown in Fig.\ref{charam} (for more details, see \cite{lectnotes}). 

The today's amplitudes in different parts of the spectrum are mostly 
determined by the thickness of the barrier $a'/a$, that is, by the duration
of time that a given mode $n$ spent under the barrier. All modes start with the 
initial $N_i=1/2$, but if the mode $n$ enters the barrier
(i.e. satisfies the condition $\lambda_i \approx c/H_i$) when the scale factor 
was $a_{i}(n)$, and leaves the barrier (i.e satisfies the condition 
$\lambda_f \approx c/H_f$) when the scale factor was $a_{f}(n)$, the final 
number of quanta in the mode will be $N_{f}(n) = \left(a_f/a_i\right)^2 =
\left(\lambda_f/\lambda_i\right)^2 = \left(H_i/H_f\right)^2 $. The very 
high-frequency modes with $n$ above the tip of the barrier, $n>n_1$, 
remain in the adibatic regime throughout their evolution, and their 
energy is being renormalised to zero. Therefore, the resulting amplitude 
spectrum quickly drops to zero at the high-frequency end.   

The power-law index of today's spectrum $h(n)$ in an arbitrary narrow 
interval of frequencies, say, between $n_a$ and $n_b$, as outlined in
Fig.\ref{charam}, is expressible in terms of the initial vacuum spectrum
$h_i(n) \propto n$ and the power-law indeces $\beta$ of the scale 
factor \cite{gr74} 
\[
a(\eta) \propto  |\eta|^{1+ \beta}
\]
which approximates exact cosmological evolution in short intervals 
of time when the left and the right sides 
of the relevant portion of the barrier $a'/a$ were formed. Today's 
spectral index of $h(n)$ in the discussed narrow interval of frequencies 
is given by 
\begin{equation}
\label{qualsp}
h(n) \propto n~n^{1+\beta_i}~n^{-(1+\beta_f)} \propto n^{1+\beta_i - \beta_f},
\end{equation}
where $\beta_i$ and $\beta_f$ refer to the left and to the 
right slopes of the barrier, respectively.

The vacuum spectrum processed only on the left 
slope of the barrier, i.e. considered at times before 
processing on the right slope of the barrier, is called the primordial 
spectrum $h_{p}(n)$:
\begin{equation}
\label{prims}
h_{p}(n) \propto n^{2+\beta_i}.
\end{equation}
For some historical and notational reasons, one and the same spectral 
index $2(2+\beta_i)$ in the power spectrum $h^2(n)$ of primordial metric 
perturbations is called $n_t$ in the case of gravitational waves 
and $n_s -1$ in the case of density perturbations (more details below,
Eq.(\ref{spectins})). We will use the notation $2(2+ \beta_i) = {\rm n} -1$. 

If the cosmological evolution $a(\eta)$ is such
that the barrier $a'/a$ is `one-sided' and has only the right slope, the 
determination of the primordial spectrum requires additional considerations.
The under-barrier amplification of the waves is still taking place 
\cite{gr74}, but the lack of the initial high-frequency regime makes the 
definition of the initial amplitudes somewhat ambiguous.

To summarise, for a given cosmological evolution $a(\eta)$, one can 
qualitatively predict today's piecewise amplitudes and slopes of the 
spectrum $h(n)$ without doing any particularly detailed calculations. 
A barrier, whose shape is more complicated than the one shown in 
Fig.\ref{charam}, would have resulted in a more complicated shape of the 
generated spectrum. On the other hand, from the measured g.w. spectrum 
one can, in principle, reconstruct cosmological evolution 
$a(\eta)$ \cite{grsol}.

We will now turn to rigorous calculations based on quantum theory.   
The traditional, but not obligatory \cite{weinbB}, approach to quantum 
theory begins with a classical Lagrangian. The  
Lagrangian for a gravitational-wave oscillator of frequency $n$, and 
for each of two polarizations $s=1,2$, has the form (\cite{gr05} and
references there):
\begin{equation}
\label{Lg2}
L_{gw} = \frac{1}{2n} \left(\frac{a}{a_0}\right)^2 
\left[\left( {\bar{h}}'\right)^2 -n^2 {\bar{h}}^2\right], 
\end{equation}
where
\begin{equation}
\label{hbar}
\bar{h} = \left(\frac{\hbar}{32 \pi^3}\right)^{1/2}\frac{\lambda_0}{l_{Pl}} h. 
\end{equation}
This Lagrangian can be derived from the total Hilbert-Einstein quadratic 
action, where both gravity and matter parts of the action are taken into 
account. The classical equation of motion derivable from the Lagrangian 
(\ref{Lg2}) in terms of the variable $h$ is Eq.(\ref{heq}),
and the equation of motion derivable in terms of the variable $\mu$, 
where $h= \mu/a$, is Eq.(\ref{meq}).

The canonical pair of position $q$ and momentum $p$ for the
oscillator can be taken as
\begin{equation}
\label{mom}
q = \bar{h}, ~~~~~~~~p = \frac{\partial L_{gw}}{\partial {\bar{h}}^{\prime}} =
\frac{1}{n}\left(\frac{a}{a_0}\right)^2 {\bar{h}}^{\prime}.
\end{equation} 
Then, the classical Hamiltonian $H_{gw} = p q' - L_{gw}$ reads
\begin{equation}
H_{gw} = \frac{n}{2}\left[ \left(\frac{a_0}{a}\right)^2 p^2 +
\left(\frac{a}{a_0}\right)^2 q^2\right].
\label{Hgwcl}
\end{equation}

We now promote $q$ and $p$ to the status of quantum-mechanical operators and 
denote them by bold-face letters. Since we are interested in the initial
conditions imposed in the early high-frequency regime, i. e. when $n \gg a'/a$,
we can write the following asymptotic expressions for the operators:
\begin{equation}
\label{qp1}
{\bf q}= \sqrt{\frac{\hbar}{2}} \frac{a_0}{a}\left[{\bf c}e^{-in(\eta-\eta_0)} 
+ {\bf c}^{\dagger}e^{in(\eta-\eta_0)} \right],
\end{equation}
\begin{equation}
\label{qp2}
{\bf p}= i\sqrt{\frac{\hbar}{2}} \frac{a}{a_0}\left[-{\bf c}
e^{-in(\eta-\eta_0)} + {\bf c}^{\dagger}e^{in(\eta-\eta_0)} \right].
\end{equation}

The commutation relationships for ${\bf q}, {\bf p}$ operators, and
for the annihilation and creation ${\bf c}, {\bf c}^{\dagger}$ operators, 
are given by
\[
\left[{\bf q},{\bf p} \right] =i\hbar, ~~~~~~~~~~~\left[ {\bf c}, 
{\bf c}^{\dagger}\right]=1.
\]
The asymptotic expression for the Hamiltonian $H_{gw}$ takes the form
\begin{equation}
{\bf H}_{gw} = \hbar~n~{\bf c}^{\dagger} {\bf c}.
\label{Hgw}
\end{equation}

Obviously, the quantum state $|0\rangle$ satisfying the condition 
\[
{\bf c}|0\rangle =0
\]
is the state of the lowest energy, i.e. the ground (vacuum) state of the 
Hamiltonian (\ref{Hgw}). At $\eta = \eta_0$ we get the relationships
\[
\langle 0|{\bf q}^2|0 \rangle = \langle 0|{\bf p}^2|0 \rangle =\frac{\hbar}{2},
~~~~~~~~\Delta {\bf q} \Delta {\bf p} =\frac{\hbar}{2}. 
\]

The root-mean-square value of ${\bf q}$ in the vacuum state $|0\rangle$
is $q_{rms} = \sqrt{\hbar/2}$. Combining this number with (\ref{hbar}) we 
derive
\begin{equation}
\label{hrmsgw}
h_{rms} = {\left(\langle 0|{\bf h}^2|0 \rangle\right)}^{1/2} =
\frac{\sqrt{2}(2 \pi)^{3/2} l_{Pl}}{\lambda_0},
\end{equation}
which agrees with the qualitative estimate (\ref{iampl}).The adopted
notations require ${\cal C} = \sqrt{16 \pi} l_{Pl}$ in the 
general expression (\ref{fourierh}).

Now we will have to discuss exact Hamiltonians. The Hamiltonian built on 
the canonical pair $\mu,~ \partial L_{gw}/\partial \mu^{\prime}$ manifestly 
illustrates the underlying pair creation process for gravitational waves. 
Specifically, we introduce 
\[
Q= \sqrt{\frac{\hbar}{8 \pi}}\frac{1}{n \sqrt{n}} \frac{1}{l_{Pl}} \mu,
~~~~~P = \frac{\partial L_{gw}}{\partial Q'} = Q' - \frac{a'}{a} Q
\]
and write the classical Hamiltonian in the form
\[
H_{gw} =\frac{1}{2}\left[ P^2 + n^2 Q^2 + \frac{a'}{a}
\left(PQ + QP \right)\right].
\]
The associated annihilation and creation operators are
\[
c = \sqrt{\frac{n}{2}} \left(Q +i \frac{P}{n}\right), ~~~~
c^{\dagger} = \sqrt{\frac{n}{2}} \left(Q - i \frac{P}{n}\right).
\]
The full quantum Hamiltonian can be written as \cite{gr93}
\begin{equation}
\label{hamgw}
{\bf H}(\eta) = n {\bf c}^{\dagger}{\bf c} + \sigma {{\bf c}^{\dagger}}^2 +
\sigma^{*} {\bf c}^2,
\end{equation}
where coupling to the external field is given by the function
$\sigma(\eta) = (i/2)(a^{\prime}/a)$. 

The interaction with the external field
can be neglected when $n \gg a'/a$, and then the first term in (\ref{hamgw}) 
dominates and represents a free oscillator. The early-time asymptotic 
expressions for the Heisenberg operators,
\[
{\bf c}(\eta) = {\bf c} e^{-in(\eta- \eta_0)}, ~~~~~
{\bf c}^{\dagger}(\eta) = {\bf c}^{\dagger} e^{in(\eta-\eta_0)},
\] 
enter into formulas (\ref{qp1}), (\ref{qp2}). 

One can note that the notion of the barrier $a'/a$ is convenient when one 
thinks of the problem in terms of the interaction Hamiltonian (\ref{hamgw}) 
and the first-order Heisenberg equations of motion, whereas 
the barrier $a''/a$ is more convenient when one thinks of the problem in 
terms of the second-order Schrodinger-like equation (\ref{meq}).

The emerging correlation between the traveling modes ${\bf n}$ and 
$-{\bf n}$, which leads to the production of standing waves referred to in the 
Introduction, is described by the 2-mode Hamiltonian (see \cite{gr93} and 
references there):
\begin{equation}
\label{hamgw2}
{\bf H}(\eta) = n {\bf c}_{\bf n}^{\dagger}{\bf c}_{\bf n} + 
n {\bf c}_{-{\bf n}}^{\dagger}{\bf c}_{-{\bf n}} + 
2\sigma {\bf c}_{\bf n}^{\dagger}{\bf c}_{-{\bf n}}^{\dagger} +
2\sigma^{*} {\bf c}_{\bf n}{\bf c}_{-{\bf n}}.
\end{equation}
This Hamiltonian can be viewed as the sum of two Hamiltonians
(\ref{hamgw}). The Hamiltonian (\ref{hamgw}) is called a 1-mode
Hamiltonian.

The defined quantum-mechanical operators and Hamiltonians, plus the 
assumption about a particular initial state of the field, fully 
determine dynamical evolution of the field, its statistical 
properties, correlation functions, and eventually the observational 
predictions for later times.

\section{\label{sec:modf}Squeezing and Power Spectrum}

The quantum-mechanical Schrodinger evolution transforms the initial
vacuum state $|0_{\bf n} \rangle |0_{-{\bf n}} \rangle$ into a 2-mode
squeezed vacuum state, which is equivalent to a pair of 1-mode squeezed
vacuum states (see \cite{gr93} and references there). In the Heisenberg
picture, the initial state of the field does not evolve, but the operators 
do, and their evolution is ultimately described by the mode functions
$\stackrel{s}{h}_n(\eta)$.

It is the variance of the oscillator's phase that is being strongly diminished
(squeezed), whereas the variance of the amplitude is being strongly increased. 
The Gaussian nature of the initial vacuum state is maintained in the course 
of the Schrodinger evolution, but the variances of phase and amplitude in the 
resulting squeezed vacuum quantum state are dramatically different. The 
parameter of squeezing and the mean number of quanta are growing all the way 
up in the amplifying regime, and stop growing only at its end \cite{gr93}. 
(The multi-quantum nature of the developing 
squeezed vacuum quantum state is behind the continuing debate over the 
`quantum-to-classical' transition, `decoherence', etc.) The phenomenon of 
squeezing allows us to treat the resulting quantum states as a stochastic 
collection of standing waves. The squeezing and the associated picture of 
standing waves is very important observationally \cite{gr93}. It leads to 
oscillatory features in the metric power spectrum and, as a consequence, 
to oscillatory features in the angular power spectrum of CMB temperature and 
polarization anisotropies. We will discuss these oscillations later. 

Having accepted the initial vacuum state $|0 \rangle$ of the gravitational 
(metric) field (\ref{fourierh}), we can calculate the variance of the field:
\begin{eqnarray}
\left<0\right|h_{ij}(\eta,{\bf x})h^{ij}(\eta,{\bf
x})\left|0\right> =
\frac{\mathcal{C}^{2}}{2\pi^{2}}\int\limits_{0}^{\infty}
~n^{2}\sum_{s=1,2}|\stackrel{s}{h}_n(\eta)|^{2}\frac{dn}{n}.
\label{meansq}
\end{eqnarray}
The quantity
\begin{eqnarray}
h^{2}(n,\eta) = \frac{\mathcal{
C}^{2}}{2\pi^{2}}n^{2}\sum_{s=1,2}|\stackrel{s}{h}_n(\eta)|^{2}
\label{gwpower}
\end{eqnarray}
gives the mean-square value of the gravitational field perturbations in a 
logarithmic interval of $n$ and is called the metric power spectrum. 
The spectrum of the root-mean-square ($rms$) 
amplitude $h(n,\eta)$ is determined by the square root of Eq.(\ref{gwpower}).

Having evolved the classical mode functions
$\stackrel{s}{h}_n(\eta)$ up to some arbitrary instant of time $\eta$ one can 
find the spectrum $h(n, \eta)$ at that instant of time. 
As mentioned before, the spectrum calculated at times when the waves of today's 
interest were in their long-wavelength regime (that is, longer than the 
Hubble radius at that times) is called the primordial spectrum.
The today's spectrum, i.e. spectrum calculated at $\eta= \eta_R$, is 
normally expressed in terms of frequency $\nu$ measured in Hz, 
$\nu =nH_0/4\pi =n \nu_H/4\pi$. The spectral $rms$-amplitude is denoted 
by $h_{rms}(\nu)$, or simply $h(\nu)$.  
The mean-square value of the gravitational-wave field in some interval of
frequencies between $\nu_1$ and $\nu_2$ is given by the integral:
\begin{equation}
\label{hmsqnu}
\left<h^2\right> = \int\limits_{\nu_1}^{\nu_2} h^2(\nu) \frac{d\nu}{\nu}.
\end{equation}
The spectral function $h^2(\nu)$ depends on frequency $\nu$,
but is dimensionless. The dimensionality of ${\rm Hz}^{-1}$ is carried by
the function $h^2(\nu)/\nu$.

For gravitational waves which are comfortably shorter than
the Hubble radius, one can also calculate the spectral gravitational-wave
energy density $\rho_{gw}(\nu)c^2$ and the total g.w. energy density in some 
interval of frequencies. These quantities are expressible in terms 
of $h^2(\nu)$:
\begin{equation}
\label{gwenden}
\rho_{gw}(\nu)= \frac{\pi}{8 G} h^2(\nu) \nu^2, ~~~~~
\rho_{gw}(\nu_1, \nu_2) = \int_{\nu_1}^{\nu_2} \rho_{gw}(\nu) \frac{d \nu}{\nu}.
\end{equation}
For the purpose of comparing a g.w. background with other sorts of matter, 
it is convenient to introduce the cosmological $\Omega_{gw}$-parameter and
its spectral value $\Omega_{gw} (\nu)$. 
As with all other sorts of radiation, this parameter is defined by 
\begin{equation}
\label{omegapar}
\Omega_{gw}(\nu) = \frac{\rho_{gw}(\nu)}{\rho_{crit}},
\end{equation}
where $\rho_{crit} = 3 H_0^2/8\pi G$. Using Eq.(\ref{gwenden}) one can also 
write \cite{lectnotes}
\begin{equation}
\label{omegafor}
\Omega_{gw}(\nu) = \frac{\pi^2}{3} h^2(\nu) \left(\frac{\nu}{\nu_H}\right)^2. 
\end{equation}
The total $\Omega_{gw}$ between frequencies $\nu_1$, $\nu_2$ is given by
\[
\Omega_{gw}(\nu_1, \nu_2) =  \int_{\nu_1}^{\nu_2} \Omega_{gw}(\nu) 
\frac{d \nu}{\nu}.
\]

One should be weary of the confusing definition
\[ 
\Omega_{gw}(\nu) =\frac{1}{\rho_{crit}} \frac{d \rho_{gw}(\nu)}{d \ln \nu}
\]
often floating in the literature. As it stands, this relationship is
incorrect. It can be made consistent with
the correct definition (\ref{omegapar}) only if one assumes that what is
being differentiated in this formula is not the spectral 
density $\rho_{gw}(\nu)$, but a logarithmic integral of this quantity in 
the limits between some fixed $\nu_1$ and a running $\nu$.

To make more precise the expected qualitative graph for $h_{rms}(n)$ 
in Fig.\ref{charam}, as well as previous theoretical graphs in Fig.4 of
Ref.\cite{grpoln}, we need to use some available observational data. We make 
the fundamental assumption that the observed CMB anisotropies are indeed 
caused by cosmological perturbations of quantum-mechanical origin. If so, 
the contribution of relic gravitational waves to the large-scale anisotropies 
should be of the same order of magnitude 
as the contribution of density perturbations (we will show this in more 
detail below). This allows us to determine the position and the slope of the 
function $h_{rms}(\nu)$ at frequencies near the Hubble frequency
$\nu_H \approx 2\times 10^{-18} Hz$ and then extrapolate the spectrum to
higher frequencies. 

\begin{figure}
\begin{center}
\includegraphics[width=15cm]{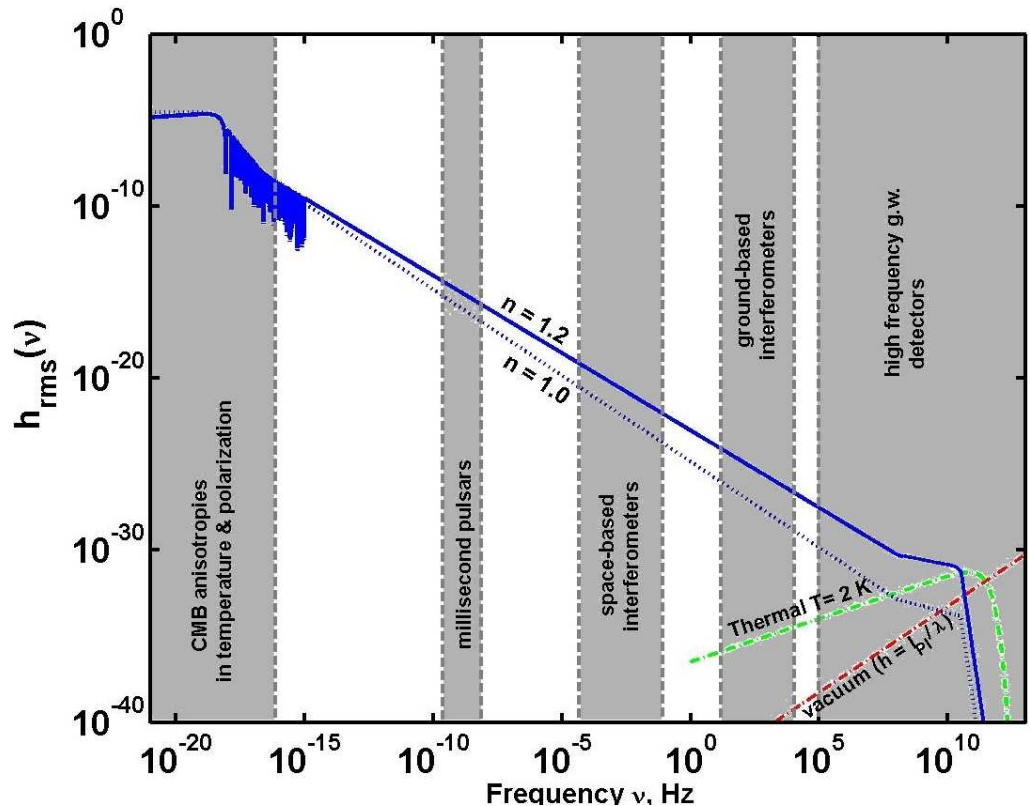}
\end{center}
\caption{The present-day spectrum for $h_{rms}(\nu)$. The solid line 
corresponds to the primordial
spectral index $\beta =-1.9$, i.e. ${\rm n} =1.2$, while the
dashed line is for $\beta= -2$, i.e. ${\rm n} =1$.}
\label{gwSpectrum2}
\end{figure}

We choose $h_{rms}(\nu_H)$ and primordial spectral index $\rm n$ in such a 
way that the lower-order CMB multipoles produced by relic gravitational 
waves are at the level of the actually observed values. Then, 
today's spectra for $h_{rms}(\nu)$ with spectral indeces ${\rm n}=1$ 
($\beta_i = -2$) and ${\rm n} =1.2$ ($\beta_i= -1.9$) are shown in 
Fig.\ref{gwSpectrum2} 
(for more details, see \cite{bgp} and \cite{lectnotes}).
At frequencies around $\nu_H$ we have 
$h_{rms}(\nu_H)\approx 10^{-5}$, so that the present-day mean number of 
quanta $N_f$ exceeds $10^{100}$. 

When extrapolating the spectrum to higher frequencies we assume that the
entire left slope of the barrier in Fig.\ref{a'a} was formed by a single 
power-law scale factor with one and the same $\beta_i =\beta$:
\begin{equation}
\label{inscf}
a(\eta) = l_o|\eta|^{1+ \beta}.
\end{equation}
This seems to be a reasonable assumption given the relative featurelessness
of the initial stage -- the overall energy density was 10 orders of magnitude 
lower than the Planckian density and was barely changing. Specifically, at the
$i$-stage, we consider two examples: $\beta = -2$ and $\beta= -1.9$. The 
power-law indeces $\beta_f$ at the matter-dominated and radiation-dominated 
stages are well known: $\beta_f=1$ and $\beta_f=0$, respectively. The effective
pressure $p$ and the energy density $\epsilon$ of matter driving the general 
power-law evolution (\ref{inscf}) with a given constant $\beta$ are related by 
the effective equation of state $p = [(1-\beta)/3(1+\beta)] \epsilon$.

As for the high-frequency part of the spectrum, it was calculated under the 
assumption that the $z$-stage of evolution shown in Fig.\ref{scalef} was 
governed by matter with a stiff equation of state $p = \epsilon$ 
($\beta_f= -(1/2)$) advocated long ago by Zeldovich. The issue of the 
back reaction of the created gravitons on the ``pump" field $a(\eta)$ 
becomes important for this sort of values of $\beta_f$ \cite{gr74}, 
\cite{zelnov}. Of course, the existence of such an interval in the past 
evolution of the very early Universe cannot be guaranteed. In any case, the 
waves with frequencies above $10^{10}$ Hz have never been in 
superadiabatic regime, they remain in the vacuum state. The renormalization
(subtraction of a ``half of the quantum" from each mode) cuts off the spectrum 
at these high frequencies. At lower frequencies, the renormalization has 
practically no effect on the spectrum.

\begin{figure}
\begin{center}
\includegraphics[width=15cm]{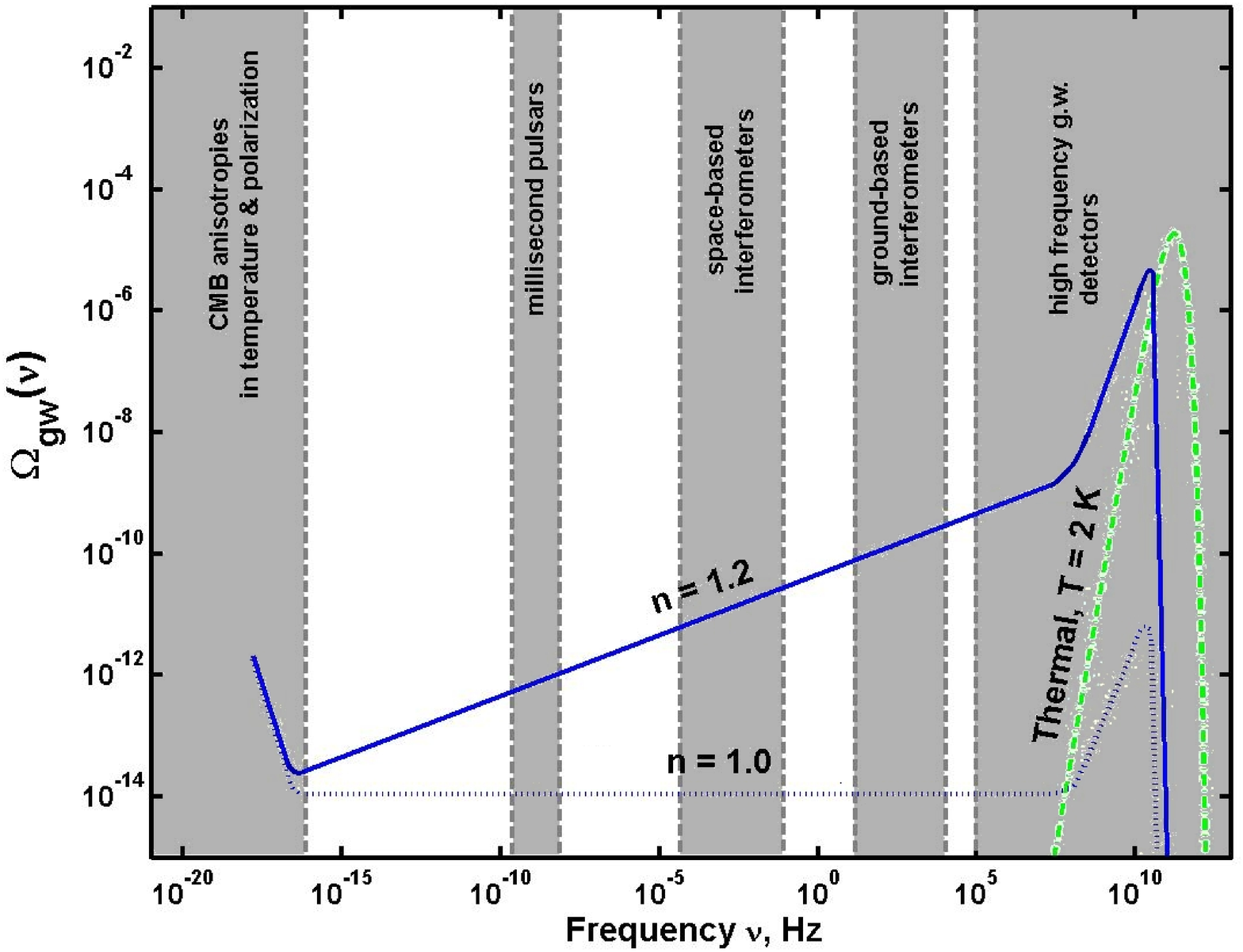}
\end{center}
\caption{The present-day spectrum for $\Omega_{gw}(\nu)$. The solid line 
corresponds to the primordial
spectral index $\beta =-1.9$, i.e. ${\rm n} =1.2$, while the
dashed line is for $\beta= -2$, i.e. ${\rm n} =1$.}
\label{OmegaSpectrum2}
\end{figure}

The spectra of $\Omega_{gw}(\nu)$, shown in Fig.\ref{OmegaSpectrum2}, 
are derived from $h_{rms}(\nu)$ according to Eq.(\ref{omegafor}). 
As was already mentioned, the substantial rise of the spectrum
at very high frequencies cannot be guaranteed, and the shown graphs 
in this area of frequencies should be regarded as the upper allowed 
limits (especially for the model with the primordial spectral index 
${\rm n} =1.2$). 

There are two comments to be made about Fig.\ref{gwSpectrum2} and 
Fig.\ref{OmegaSpectrum2}. First, the phenomenon of squeezing and standing 
waves production is reflected in the oscillations
of the metric power spectrum and $\Omega$-spectrum as functions of 
frequency $\nu$. The first few cycles of these oscillations are shown 
in the graph of $h_{rms}(\nu)$. In the CMB sections of the paper, we will 
show how these oscillations that existed in the recombination era translate 
into peaks and dips of the angular power spectra for CMB temperature and 
polarization anisotropies observed today. 

Second, since the primordial spectrum of quantum-mechanically generated
perturbations has the form (see Eq.(\ref{prims})):
\begin{equation}
\label{primsp}
h^{2}(n) \propto n^{2(\beta+2)},
\end{equation}
theoretical considerations suggest that the preferred range of values for the 
primordial spectral index ${\rm n}$ is ${\rm n} > 1$, i.e. preferred spectra
are `blue'. Indeed, for the opposite range ${\rm n} < 1$, the primordial 
spectra would be `red', and the integral (\ref{meansq}) would be 
power-law divergent at the lower limit. To avoid the infra-red divergency 
one would need to make extra assumptions about bending the spectrum down in 
the region of very small wavenumbers $n$. In its turn, a primordial spectral 
index ${\rm n} > 1$ requires that the power-law index $\beta$ of the
initial scale factor $a(\eta)$ should be $\beta > -2$. This is because 
$\beta$ and ${\rm n}$ are related  by ${\rm n} = 2 \beta +5$. The 
value ${\rm n} =1$ ($\beta = -2$) describes the spectrum known as a 
flat, or Harrison-Zeldovich-Peebles, or scale-invariant spectrum. It was
originally proposed in the context of density perturbations. 
The spectral index ${\rm n} =1$ marks the beginning of the trouble -- 
the metric variance (\ref{meansq}) is logarithmically divergent at the 
lower limit.

\section{\label{sec:denp}Density Perturbations}

To discover relic gravitational waves in CMB anisotropies we have to 
distinguish their effects from other possible sources of CMB anisotropies,
and first of all from density perturbations. In comparison with gravitational
waves, the possibility of quantum-mechanical generation of density 
perturbations requires an extra hypothesis, namely, the appropriate 
parametric coupling of the matter field to external gravity. But we 
can assume that this hypothesis was satisfied by, for example, a lucky 
version of the scalar field. Can the amplitudes of quantum-mechanically 
generated density perturbations be many orders of magnitude larger than 
the amplitudes of relic gravitational waves ? This would make the 
search for relic gravitational waves practically hopeless. We will show 
that this can never happen. Density perturbations can compete with relic 
gravitational waves in producing the large-scale CMB anisotropies, but can 
never be overwhelming.

We are mostly interested in the gravitational field (metric) sector
of density perturbations. Indeed, it is perturbations in the gravitational 
field that survived numerous transformations of the matter content of the 
Universe. Primordial matter (for example, scalar field) has decayed,
together with its own perturbations, long ago. But it is primordial 
metric perturbations that have been inhereted by density perturbations
at radiation-dominated and matter-dominated stages. 

As a driving `engine' for the very early Universe, it is common  
to consider the so-called minimally-coupled scalar field 
$\varphi(\eta,{\bf x})$, with the energy-momentum tensor
\begin{equation}
\label{Tmn}
T_{\mu \nu} = \varphi_{,\mu} \varphi_{,\nu} - g_{\mu \nu} \left[
\frac{1}{2}g^{\alpha \beta} \varphi_{,\alpha} \varphi_{,\beta} + 
V(\varphi) \right].
\end{equation}
The polarization structure (\ref{ptensors4}) of density perturbations
allows us to write the ${\bf n}$-mode metric perturbation:
\begin{equation}
\label{hijd}
h_{ij}= h(\eta) Q \delta_{ij} + h_l(\eta) n^{-2} Q_{,ij},
\end{equation}
where $Q = e^{\pm i{\bf n}\cdot{\bf x}}$. Two polarization  
amplitudes $h(\eta)$ and $h_l(\eta)$ are accompanied by the third
unknown function - the amplitude 
$\varphi_1(\eta)$ of the scalar field perturbation:
\[
\varphi(\eta, {\bf x}) = \varphi_0(\eta) + \varphi_1(\eta) Q. 
\]

The perturbed Einstein equations allow us to find all three unknown functions.
The important fact is that, for any $V(\varphi)$, there exists only one 
second-order differential equation, of the same structure as Eq.(\ref{meq}), 
that needs to be solved \cite{gr94}:
\begin{equation}
\label{dpeq}
\mu^{\prime\prime} + \mu \left[n^2 - \frac{(a \sqrt{\gamma})^{\prime\prime}}
{a \sqrt{\gamma}}\right] = 0.
\end{equation}
This equation coincides with the gravitational wave equation (\ref{meq}) if 
one makes there the replacement
\begin{equation}
\label{newsf}
a(\eta) \rightarrow a(\eta) \sqrt{\gamma(\eta)}.
\end{equation}
We will call function $\mu$ satisfying
Eq.(\ref{dpeq}) $\mu_S$, and that satisfying Eq.(\ref{meq}) $\mu_T$.

The function $\gamma(\eta)$ is defined by 
\[
\gamma(\eta) = 1 + \left(\frac{a}{a^{\prime}} \right) ^{\prime} =
- \frac{c}{a} \frac{H^{\prime}}{H^2}.
\]
For power-law scale factors (\ref{inscf}), this 
function reduces to a set of constants,
\[
\gamma = \frac{2+\beta}{1+\beta}.
\]
The constant $\gamma$ degenerates to zero in the limit of the expansion
law with $\beta = -2$, that is, in the limit of the gravitational
pump field (an interval of deSitter evolution) which is responsible for 
the generation of cosmological perturbations with flat primordial 
spectrum ${\rm n} =1$. 

In terms of $t$-time the function $\gamma$ is
\[
\gamma(t) =  - \frac{\dot H}{H^2}.
\]
It is also related to the cosmological decelaration parameter 
$q(t)=-a\ddot{a}/{\dot{a}}^2$: $\gamma(t) = 1 + q(t)$. The unperturbed 
Einstein equations allow us to write
\begin{equation}
\label{phipr}
\kappa \left({\varphi_0}^{\prime}\right)^2 =2 \left(\frac{a'}{a}\right)^2
\gamma(\eta),
\end{equation}
or, equivalently,
\begin{equation}
\label{phidot}
\frac{\dot{\varphi}_0}{H} =\sqrt{\frac{2}{\kappa}}\sqrt{\gamma(t)},
\end{equation}
where $\kappa = 8\pi G/c^4$. The function $\gamma(t)$ is sometimes 
denoted in the literature by $\epsilon(t)$.

As soon at the appropriate solution to Eq.(\ref{dpeq}) is found, all
unknown functions are easily calculable:
\begin{equation}
\label{hsol}
h(\eta) = \frac {1}{c} H(\eta)\left[ \int_{\eta_0}^{\eta} \mu \sqrt{\gamma}
d\eta +C_i\right],
\end{equation}
\begin{equation}
\label{hlsol}
{h_l}^{\prime}(\eta)= \frac{a}{a'} \left[h''- \frac{H''}{H'} h' +n^2 h 
\right],
\end{equation}
\begin{equation}
\label{phisol}
\varphi_1(\eta) = \frac{\sqrt \gamma}{\sqrt{2 \kappa}} \left[\frac{\mu}
{a \sqrt \gamma} - h \right].
\end{equation}

The arbitrary constant $C_i$ in Eq.(\ref{hsol}) reflects the remaining 
coordinate freedom within the class of synchronous coordinate 
systems (\ref{FRWmetric}). Indeed, a small coordinate 
transformation \cite{gr94}
\begin{equation}
\label{etatr}
\overline{\eta} = \eta - \frac{C}{2a} Q
\end{equation}
generates a `gauge transformation'
\begin{equation}
\label{hgauge}
\overline{h}(\eta) = h(\eta) + C \frac{a'}{a^2} = h(\eta) + C \frac{H(\eta)}{c}.
\end{equation}
The term with constant $C$ is automatically present in the properly
written general solution (\ref{hsol}) for $h(\eta)$.

By combining the transformation (\ref{hgauge}) with its time-derivative, 
one can build a quantity which does not contain the constant $C$ and 
therefore is a `gauge-invariant' quantity with respect to transformations 
preserving synchronous coordinates. Specifically, one can check that 
\[
h - h'\frac{H}{H'} = \overline{h} - \overline{h}' \frac{H}{H'}.
\] 
Denoting 
this quantity by $\zeta$, and taking into account Eq.(\ref{hsol}), we write
\begin{equation}
\label{zeta}
\zeta \equiv h- h'\frac{H}{H'} = \frac{\mu_S}{a \sqrt \gamma}.
\end{equation}

In terms of the variable $\zeta(\eta)$ the fundamental equation (\ref{dpeq})
takes the form
\begin{equation}
\label{zheq}
\zeta^{\prime \prime} +2\frac{(a \sqrt \gamma)^{\prime}}{a \sqrt \gamma} 
\zeta ^{\prime}+n^2 \zeta=0.
\end{equation}
Not surprisingly, this equation coincides with the gravitational-wave 
equation (\ref{heq}), if
one replaces the gravitational-wave function $h$ with $\zeta$, and $a$ 
with $a\sqrt{\gamma}$. For density perturbations, the metric variable $\zeta$
is the physically relevant quantity that plays the same role as the metric 
variable $h$ for gravitational-wave perturbations. 

The fact that we are working
with scalar gravitational waves supported by scalar field fluctuations, 
instead of normal tensor gravitational waves, has
boiled down to the necessity of a single modification: the 
substitution (\ref{newsf}) in the gravitational-wave equations.

It is appropriate to say a few words about cosmological gauge 
transformations in general. Their origin is related to the notion of 
Lie transport performed on a manifold covered by some coordinates 
$x^{\alpha}$ \cite{Lie}. Lie transport is being carried out along a 
given vector field $\xi^{\mu}(x^\alpha)$. This is a quite formal 
construction which respects only the transformation properties of fields 
defined on the manifold. These fields are not required to satisfy any 
physical equations. An infinitesimal Lie transport changes the field by 
an amount equal to its Lie derivative. This change can be viewed as a rule
by which new values of the field are assigned to the same point $x^{\alpha}$. 
It is only with some reservation that we borrow the name gauge 
transformation from physical field theories and apply it to this, always 
valid, mathematical procedure.

In gravitational applications, the vector field $\xi^\mu(x^{\alpha})$ is 
usually associated with an infinitesimal coordinate transformation 
$\overline{x}^\mu = x^\mu - \xi^{\mu}(x^\alpha)$ \cite{LL}. Then, for example, 
an infinitesimal increment of the metric tensor $g_{\mu \nu}(x^\alpha)$ is
given by the Lie derivative of $g_{\mu \nu}$. It can be written as
\[
\delta g_{\mu \nu}(x^\alpha) = \xi_{\mu ; \nu} + \xi_{\nu ; \mu}.
\]
A particular case of these transformations is represented by our 
Eqs.(\ref{etatr}, \ref{hgauge}). 

Since we operate with only four components of an arbitrary vector $\xi^{\mu}$
and, potentially, with many fields transforming with the same $\xi^{\mu}$, 
one can build some combinations of these transformed fields, 
in which the functions $\xi^{\mu}$ cancel out. 
Therefore, these combinations become gauge-invariant
quantities. An example is given by our Eq.(\ref{zeta}). 

Certainly, it is an 
exaggaration to claim that ``only gauge-invariant quantities have any 
inherent physical meaning". It is as if you were told that where specifically
you are going in your car has no physical meaning, because the components 
of velocity are coordinate-dependent, and it is only the readings of your 
speedometer that have an inherent physical meaning, because they are 
coordinate-independent. In any case, there exists an infinite number of 
gauge-invariant quantities. For example, a product of a gauge-invariant 
quantity with any `background' function produces a new gauge-invariant 
quantity. The fact that a quantity is gauge-invariant does not answer 
the question of its physical interpretation. Nevertheless, it is useful to 
know gauge-invariant quantities. 

Had we started from arbitrary coordinates, the perturbed metric
(\ref{FRWmetric}) would have contained two more unknown functions,
namely $A(\eta)$ and $B(\eta)$, \cite{bard}
\[
g_{\eta \eta}= -a^2(\eta)(1+ 2 A(\eta) Q), ~~~~~g_{\eta i} = a^2(\eta)
B(\eta) \frac{1}{n} Q_{,i}.
\]
The $\eta$-transformation generalizing our Eq.(\ref{etatr}) would 
read
\begin{equation}
\label{etatr1}
\overline{\eta} = \eta + T(\eta) Q,
\end{equation}
where $T(\eta)$ is an arbitrary function of $\eta$. The perturbed metric 
components $h$ and $A$  would transform as follows (any possible 
$x^i$-transformations do not participate in these relationships):
\begin{equation}
\label{htrans1}
\overline{h} = h - 2 \frac{a'}{a} T,~~~~ \overline{A}=A -T' -\frac{a'}{a} T.
\end{equation}

The functions $T$ and $T'$ cancel out in the gauge-invariant 
metric combination
\[
\zeta_g = h- \frac{H}{H'} h' - \frac{2A}{\gamma}.
\]
Obviously, $\zeta_g$ reduces to $\zeta$ for transformations (\ref{etatr}) 
preserving synchronous conditions, that is, when $A= \overline{A}= 0$. The 
previously introduced \cite{BST} quantity $\zeta_{BST}$, 
where $BST$ stands for Bardeen, Steinhardt, Turner, can also be reduced,
after some work, to our $\zeta$, up to a coefficient $-(1/2)$.

Since we assume that in addition to the tensor metric field $g_{\mu\nu}$, 
a scalar field 
\[
\varphi(\eta, {\bf x}) = \varphi_0(\eta) + \delta\varphi(\eta) Q 
\]
is also defined on the manifold, we can write down its gauge 
transformation under the action of Eq.(\ref{etatr1}):
\begin{equation}
\label{dft}
\overline{\delta\varphi}= \delta\varphi +\varphi_{0}^{\prime} T.
\end{equation}
The arbitrary function $T(\eta)$ cancels out in the gauge-invariant combination
\begin{equation}
\label{V}
V(\eta) = \delta \varphi +\frac{1}{2} \frac{a}{a'} \varphi_{0}' h.
\end{equation}

Although the combination (\ref{V}) is often quoted in the 
literature, this object is something like a `half of a horse plus half 
of a cow'. It combines physically separate quantities -- one from metric 
another from matter -- and its physical interpretation is obscure. 
However, using the unperturbed Einstein equation (\ref{phipr})
and solution (\ref{phisol}) for $\delta\varphi = \varphi_1$ one can
show that $V$ reduces to the gauge-invariant metric
perturbation $\zeta$ times a background-dependent factor:
\[
V= \frac{1}{\sqrt{2\kappa}} \sqrt{\gamma} \zeta.
\]
The quantity $\sqrt{2\kappa} a V$ is called Chibisov, Lukash,
Mukhanov, Sasaki \cite{luk, chm, sas} variable $u_{CLMS}$:
\begin{equation}
\label{uclm}
u_{CLMS} = a \sqrt{\gamma} \zeta.
\end{equation}
In our approach, this variable is simply the function $\mu_S$ satisfying 
Eq.(\ref{dpeq}), modeled on Eq.(\ref{meq}). The factor 
$\sqrt{\gamma}$ in Eq.(\ref{uclm}) will be a matter of great attention when
we come to the discussion of quantization procedures. 

Returning to the function $\zeta(\eta)$, one can notice that this metric
amplitude is practically constant in the regime when $n^2$ is much smaller 
than the effective potential $(a\sqrt\gamma)''/a\sqrt\gamma$.
This behaviour is similar to the constancy of the gravitational wave 
amplitude $h=\mu_T/a$ throughout the long-wavelength regime. Indeed, in 
full analogy with the long-wavelength solution (\ref{lwapp}), 
the general solution to Eq.(\ref{dpeq}) in this regime reads 
\[
\frac{\mu_S}{a \sqrt{\gamma}} = C_1 +C_2 \int \frac{d\eta}{(a \sqrt{\gamma})^2}.\]
For usually considered expanding cosmologies, the term with
constant $C_2$ is decreasing, and therefore the dominant solution is
$h \approx C_1$ for gravitational waves and $\zeta \approx C_1$ for density
perturbations. The constancy of $\zeta$ allows
one to easily estimate the value of metric perturbations at much later times, 
at the radiation-dominated and matter-dominated stages, as soon as one knows 
the initial value of $\zeta$.

The constancy of $\zeta$ (when one can neglect the term with $C_2$) is 
sometimes called a conservation law. This association is incorrect.
Genuine conservation laws reflect symmetries of the system, and conserved
quantities are constants independently of the initial conditions. For example,
the energy of a free oscillator is a constant independently of initial 
positions and velocities. In our problem, the genuine conservation law 
for $\zeta$ would look like an empty statement $0=0$.

\section{\label{sec:qdenp}Quantization of Density Perturbations}

A quantum system is defined by its Hamiltonian. Whether or not the
Hamiltonian follows from some assumed classical Lagrangian should not be 
a question of major concern \cite{weinbB}. From this point of view, the 
quantization of density perturbations, similarly to the quantization of
gravitational waves (\ref{hamgw}), is defined by the Hamiltonian \cite{gr94}
\begin{equation}
\label{hamdp}
{\bf H}(\eta) = n {\bf c}^{\dagger}{\bf c} + \sigma {{\bf c}^{\dagger}}^2 +
\sigma^{*} {\bf c}^2.
\end{equation}
The pair creation of scalar gravitational waves $\zeta$ by the external 
pump field is regulated by the coupling function
$\sigma(\eta) = (i/2)[(a \sqrt{\gamma})^{\prime}/(a \sqrt{\gamma})]$.
However, a more traditional approach begins with 
a Lagrangian, and here too, the Hamiltonian (\ref{hamdp}) can be derived 
from quadratic perturbation terms in the total Hilbert-Einstein action, 
where both gravity and matter (scalar field) are taken into account.

The Lagrangian for an $n$-mode of density perturbations, after some 
transformations of the total Hilbert-Einstein quadratic action, can be 
written in the form \cite{gr05}
\begin{equation}
\label{Ld2}
L_{dp} = \frac{1}{2 n}\left(\frac{a \sqrt \gamma}
{a_0 \sqrt{\gamma_0}} \right)^2\left[\left({\bar{\zeta}}^{\prime}\right)^2 -
n^2 {\bar{\zeta}}^2\right], 
\end{equation}
where 
\begin{equation}
\label{zbar}
\bar{\zeta} = \left(\frac{\hbar}{32 \pi^3}\right)^{1/2}
\frac{\lambda_0}{l_{Pl}} \zeta,
\end{equation}
and $a_0$, $\gamma_0$ are values of the functions $a(\eta)$, $\gamma(\eta)$
at $\eta=\eta_0$ where the initial conditions are being set. The 
Euler-Lagrange equations derivable from this Lagrangian in terms of
$\zeta$ are Eq.(\ref{zheq}), and in terms of $\mu_S$ -- Eq.(\ref{dpeq}).   

Obviously, the Lagrangian (\ref{Ld2}) for $\zeta$ coincides with the 
gravitational wave Lagrangian (\ref{Lg2}) for $h$ after the replacement
of the factor $a/a_0$ with the factor $a \sqrt{\gamma}/a_0 \sqrt{\gamma_0}$.
Starting from the Lagrangian (\ref{Ld2}) and building on the canonical
pair $\mu_S,~ \partial L_{dp}/\partial \mu_S^{\prime}$ one can derive the
Hamiltonian (\ref{hamdp}) by doing exactly the same steps that have led us
from the Lagrangian (\ref{Lg2}) to the Hamiltonian (\ref{hamgw}).

It should be noted that the total classical Lagrangian (\ref{Ld2}) admits, 
as always, some freedom of modifications without affecting the 
Euler-Lagrange equations. 
In particular, the derived equations of motion (\ref{zheq}), (\ref{dpeq}) 
will remain exactly the same, if one changes the Lagrangian (\ref{Ld2}) to a 
new one by multiplying (\ref{Ld2}) with a constant, for example with a 
constant $\gamma_0$:
\begin{equation}
\label{Ld2new}
L_{dp(new)} = \frac{1}{2 n}\left(\frac{a \sqrt \gamma}
{a_0 \sqrt{\gamma_0}} \right)^2 \gamma_0 \left[\left({\bar{\zeta}}^{\prime}
\right)^2 -n^2 {\bar{\zeta}}^2\right] = \frac{1}{2 n}\left(\frac{a}
{a_0} \right)^2 \gamma \left[\left({\bar{\zeta}}^{\prime}
\right)^2 -n^2 {\bar{\zeta}}^2\right]. 
\end{equation}
This new Lagrangian degenerates to zero in the limit $\gamma \rightarrow 0$. 
We will discuss later the subtleties in quantum theory that arise after such 
a modification of the Lagrangian.

Similarly to gravitational waves, we impose initial conditions in the early
regime of a free oscillator, that is, when 
$n \gg (a \sqrt{\gamma})'/(a \sqrt{\gamma})$. We choose the canonical pair
\begin{equation}
\label{momdp}
q = \bar{\zeta}, ~~~~~~~~p = \frac{\partial L_{dp}}{\partial 
{\bar{\zeta}}^{\prime}} = \frac{1}{n}\left(\frac{a \sqrt{\gamma}}
{a_0 \sqrt{\gamma_0}}\right)^2 {\bar{\zeta}}^{\prime},
\end{equation} 
and write the asymptotic expressions for the operators:
\begin{equation}
\label{qp1d1}
{\bf q}= \sqrt{\frac{\hbar}{2}} \frac{a_0 \sqrt{\gamma_0}}{a \sqrt{\gamma}}
\left[{\bf c}e^{-in(\eta-\eta_0)} 
+ {\bf c}^{\dagger}e^{in(\eta-\eta_0)} \right],
\end{equation}
\begin{equation}
\label{qp2d2}
{\bf p}= i\sqrt{\frac{\hbar}{2}} \frac{a \sqrt{\gamma}}{a_0 \sqrt{\gamma_0}}
\left[-{\bf c}
e^{-in(\eta-\eta_0)} + {\bf c}^{\dagger}e^{in(\eta-\eta_0)} \right],
\end{equation}
with
\begin{equation}
\label{comrdp}
\left[{\bf q},{\bf p} \right] =i\hbar, ~~~~~~~~~~~\left[ {\bf c}, 
{\bf c}^{\dagger}\right]=1.
\end{equation}

Obviously, a quantum state satisfying the condition
\begin{equation}
\label{vacz}
{\bf c}|0 \rangle =0
\end{equation}
is the ground (vacuum) state of the Hamiltonian (\ref{hamdp}).
Calculating the mean square values of ${\bf q}$ and its 
canonically conjugate momentum ${\bf p}$, we find 
\[
\langle 0|{\bf q}^2|0 \rangle = \langle 0|{\bf p}^2|0 \rangle =\frac{\hbar}{2},
~~~~~~~~\Delta {\bf q} \Delta {\bf p} =\frac{\hbar}{2}.
\]
Returning to $\zeta$ from $\bar{\zeta}$ according to Eq.(\ref{zbar}), we find
\begin{equation}
\label{zrmsgw}
\zeta_{rms} = {\left(\langle 0|{\bf \zeta}^2|0 \rangle\right)}^{1/2} =
\frac{\sqrt{2}(2 \pi)^{3/2} l_{Pl}}{\lambda_0},
\end{equation}
that is, exactly the same value as the initial amplitudes (\ref{hrmsgw}) for 
each of two polarization components of gravitational waves.

Extrapolating the initial time $\eta_0$ up to the boundary between the
adiabatic and superadiabatic regimes at $\eta =\eta_i$, we derive the 
evaluation $\zeta_{rms} \sim l_{Pl}/ \lambda_i$. This
evaluation, plus the constancy of the quantity $\zeta$
throughout the long-wavelength regime, is the foundation of the result
according to which the final (at the end of the long-wavelength regime) 
metric amplitudes of gravitational waves and density perturbations 
should be roughly equal to each other \cite{gr94}. 

The primordial $\zeta$-spectrum has the same form as the primordial 
gravitational-wave spectrum in Eq.(\ref{primsp}):
\begin{equation}
\label{primsps}
{\zeta}^{2}(n) \propto n^{2(\beta+2)}.
\end{equation}
In this approximation, the spectral indeces are equal:
\begin{equation}
\label{spectins}
n_{s} - 1 = n_t = 2(\beta + 2) \equiv {\rm n} - 1.
\end{equation}
Scalar fields (\ref{Tmn}) can support cosmological scale factors with 
$\beta$ in the interval $-1 \leq \beta$, $\beta \leq -2$. The ratio of 
(\ref{primsp}) to (\ref{primsps}) is approximately 1 for all spectral 
indeces near and including $\beta =-2$ (${\rm n} =1$).

Having strictly defined the dynamical equations and the initial 
(quantum ground state) conditions, one can calculate from 
formula (\ref{gwpower}) the exact power spectrum of metric perturbations 
associated with density perturbations. In this formula, two polarization 
components $\stackrel{s}{h}_n(\eta)$ are now determined by the conventions  
(\ref{ptensors4}), (\ref{hijd}), and the constant ${\cal C}$ is 
${\cal C} = \sqrt{24 \pi} l_{Pl}$. 

The employed approximations cannot guarantee that in 
the real Universe the coefficients in Eq.(\ref{primsp}) 
and Eq.(\ref{primsps}) should be exactly equal to each other. 
But there is no reason for them to be different by 
more than a numerical factor of order 1. Therefore, the lower 
order CMB multipoles, induced primarily by metric perturbations which 
are still in the long-wavelength regime, should be approximately at the 
equal numerical levels for, both, density perturbations and gravitational 
waves.  

As we shall now see the inflationary theory differs from the described 
quantum theory of density perturbations by many orders of magnitude, let alone
numerical coefficients of order 1. The discrepancy becomes infinitely large
in the limit of the observationally preferred spectral 
index ${\rm n} =1$ ($\beta = -2$, $\gamma =0$).

\section{\label{sec:infldenp}What Inflationary Theory Says About 
Density Perturbations, and What Should Be Said About Inflationary Theory}

For many years, inflationists keep insisting that in contrast to the generation 
of gravitational waves \cite{gr74}, which begins with the amplitude 
$h \approx l_{Pl}/\lambda$ (``half of the quantum in the mode") and finishes
with the amplitude $h \approx l_{Pl}/(c/H) \approx H/H_{Pl}$ and 
power $h^2 \approx (H/H_{Pl})^2$ (see \cite{gr74} and 
Secs. \ref{sec:quantgw}, \ref{sec:modf}), the generation of primordial 
density perturbations (i.e. scalar metric perturbations) is more efficient 
by many orders of magnitude. Contrary to the calculations reviewed in 
Sec. \ref{sec:qdenp}, inflationary theory claims that in a cosmological 
model with the same $H$, the resulting amplitude and power of the quantity 
$\zeta$ should contain a huge extra factor, tending to infinity in the limit 
of the standard de Sitter inflation.

Traditionally, the claimed inflationary derivation of density 
perturbations goes along the following lines. One starts from the ground
state quantum fluctuations taken from a different theory, namely from a 
theory of a free test scalar field, where 
metric perturbations are ignored altogether, and writes 
\[
\delta \varphi~\approx \frac{H}{2\pi}.
\]
Then, from the gauge transformation (\ref{dft}), assuming that the l.h.s. is 
equal to zero, one finds the characteristic time interval $\delta t$ ``to 
the end of inflation" and puts into $\delta t$ the estimate of 
$\delta \varphi$ from the above-mentioned `quantum' evaluation:
\[
\delta t~\approx \frac{\delta \varphi}{{\dot{\varphi}}_0} \approx \frac{H}
{2 \pi{\dot{\varphi}}_0}. 
\]
Then, one declares that the dimensionless ratio $\delta t/H^{-1}$ is what
determines the dimensionless metric and density variation amplitudes in the
post-inflationary universe. And this is being presented as 
the ``famous result" of inflationary theory: 
\[
\delta_H \approx \frac{\delta \rho}{\rho} \approx 
\frac{H^2}{2\pi {\dot{\varphi}}_0}= \frac{H/2\pi}{{\dot{\varphi}}_0/H}.
\]

This ``inflationary mechanism" of generation of density perturbations
is claimed to have been confirmed in numerous papers and 
books\footnote{All inflationary quotations in the present 
article are taken from real publications. However, I do not see much sense 
in giving precise references. I am confronting here something like a culture, 
distributed over hundreds of publications, rather than a persistent 
confusion of a few authors. This situation was once 
qualified \cite{unruh} as ``the controversy between Grishchuk and the rest 
of the community".}.
The most dramatic feature of the claimed inflationary result is the factor 
${\dot{\varphi}}_0/H$ in the denominator of the final expression. I call it
a `zero in the denominator' factor \cite{gr05}. According to Eq.(\ref{phidot})
this factor is $\sqrt \gamma$; in the literature, this factor appears in 
several equivalent incarnations, including such combinations 
as~~ $V_{,\varphi}/V$,~~ $1+ p/{\rho c^2}\equiv 1+w$,~~ $\epsilon$,~~ where 
$\epsilon \equiv \gamma \equiv - {\dot H}/{H^2}$,  and so on.

The early Universe Hubble parameter $H$ featuring in the numerator of the 
final expression cannot be too small because $H(t)$ in scalar field driven
cosmologies is a decreasing, or
at most constant, function of time. The parameter $H$ in the very early 
Universe cannot be smaller than, say, the $H$ in the era of primordial 
nucleosynthesis. So, 
the numerator of the ``famous result" cannot be zero, but the denominator 
can, at $\gamma =0$. Therefore, the ``famous result" prescribes the 
arbitrarily large numerical values to the amplitudes of density perturbations 
in the limit of the de Sitter evolution $\gamma =0$ ($\beta = -2$).
One has to be reminded that it is this gravitational pump field that 
generates gravitational waves and other perturbations 
with primordial spectrum ${\rm n} =1$, advocated long ago on theoretical 
grounds by Harrison, Zeldovich, Peebles, and which is in the vicinity of the 
primordial spectral shape currently believed to be preferred observationally.
 
The inflation 
theory claims to have predicted a ``nearly" scale-invariant spectrum of 
density perturbations because on the strictly scale-invariant 
spectrum the predicted amplitudes blow up to infinity.

More recent literature operates with the `curvature perturbation
${\cal R}$', equivalent to our $\zeta$ from Eq.(\ref{zeta}). A typical
quotation states: ``The amplitude of the resulting scalar curvature 
perturbation is given by 
\[
\langle {\cal R}^2 \rangle^{1/2} = \left(\frac{H}{\dot{\phi}}\right)
\langle \delta \phi^2 \rangle^{1/2} ,
\]
where $H$ is the Hubble parameter, $\dot \phi$ is the time derivative of 
the inflaton $\phi$, and $\delta \phi$ is the inflaton fluctuation on a
spatially flat hypersurface. The quantum expectation value of the inflaton 
fluctuations on super-horizon scales in the de Sitter space-time is
\[
\langle {\delta \phi}^2 \rangle = \left(\frac{H}{2 \pi}\right)^2."
\]

Here again one is invited to believe that the ground state quantum 
fluctuations in one theory (without metric perturbations) are responsible 
for the appearance of arbitrarily large (factor $\gamma$ in the denominator
of the ${\cal R}$ power spectrum) gravitational field fluctuations in 
another theory. Inflationists are keen to make statements about the
resulting gauge-invariant curvature perturbations without having included
metric perturbations in the initial conditions. The infinitely large 
curvature perturbation ${\cal R}$ is supposed 
to occur at $\gamma=0$, that is, exactly in the de Sitter model from which 
all the `quantum' reasoning about scalar field fluctuations has started. 

According to inflationary views on the generating process, the amount of 
the created scalar particles-perturbations is regulated not by the strength 
of the external gravitational `pump' field (basically, space-time curvature
in the early Universe) but by the closeness of the metric to a de Sitter
one. In a sequence of space-times with very modest and approximately equal 
values of $H$ you are supposed to be capable of generating arbitrarily large 
amplitudes of the scalar metric perturbation by simply going to smaller and 
smaller values of $\dot{H}$.  

Since the quantity ${\cal R}$ (as well as the quantity $\zeta$ and the 
gravitational wave function $h$) oscillates with a slowly decreasing 
amplitude in the initial short-wavelength regime (see Eq.(\ref{zheq})) and 
remains constant during the subsequent long-wavelength regime, the 
arbitrarily large amplitude of the resulting ``inflation-predicted"  
scalar curvature perturbation ${\cal R}$ must have been implanted from the 
very beginning, i.e. from the times in the high-frequency regime of evolution, 
when the intial quantum state for density perturbations was defined. It is 
important to remember that for many years inflationists claimed that the 
reason for the huge difference between the resulting scalar and tensor 
perurbations was the ``big amplification during reheating" experienced by the 
long-wavelength scalar metric perturbations. These days, the explanation via 
the ``big amplification during reheating" is not even mentioned. These days, 
the most sophisticated inflationary texts put forward, as the foundation for 
their belief in arbitrarily large resulting scalar metric perturbations 
${\cal R}$ and $\zeta$, the ``Bunch-Davies vacuum", i.e. a concept from the 
theory of a test scalar field, where the gravitational field (metric) 
perturbations are absent altogether. (Moreover, the Bunch-Davies vacuum was 
originally introduced as a de Sitter invariant state prohibiting the particle 
production by definition.) The absurd proposition of inflationists with regard 
to the density perturbations is sometimes called a ``classic result", and it 
is widely used for derivation of further incorrect conclusions in theory and 
data analysis.

The most common, and most damaging, inflationary claim is
that the amplitudes of relic gravitational waves must be
``suppressed", ``sub-dominant", ``negligibly small" in comparison with
primordial density perturbations, especially for models with 
$\gamma \rightarrow 0$. Having arrived at a 
divergent formula for density perturbations, i.e. with a `zero in the 
denominator' factor, inflationists compose the ratio of the 
gravitational-wave power spectrum ${\cal P}_T$ to the derived divergent scalar 
metric power spectrum ${\cal P}_{\zeta}$ (the so-called `tensor-to-scalar 
ratio' $r \equiv {\cal P}_T/{\cal P}_{\zeta}$) and write it as
\begin{equation}
\label{tsr}
r= 16 \epsilon = - 8 n_t.
\end{equation}
The inflationary theory binds the amplitude of scalar perturbations
with the spectral index, and makes the absurd prediction of arbitrarily 
large amplitudes of density perturbations in the limit of models with 
$\epsilon \equiv \gamma =0$ ($n_s =1$, $n_t =0$). But 
the inflationary ``consistency relation" (\ref{tsr}) 
encourages and misleads one to believe that everything is perfect -- as if 
it were the amount of gravitational waves that must go to zero in 
this limit. To make the wrong theory look acceptable, inflationary model 
builders keep 
${\cal P}_{\zeta}$ fixed at the observationally required level of $10^{-9}$ 
or so, and move $H/H_{Pl}$ down whenever $\epsilon$ goes to zero in the 
inflationary divergent formula
\[ 
{\cal P}_{\zeta} \approx \frac{1}{\epsilon}\left(\frac{H}{H_{Pl}}\right)^2,
\] 
thus making the amount of relic gravitational waves arbitrarily small.

It is instructive to see how the problem of initial conditions is delt
with by S. Weinberg \cite{weinI}. The author operates with equations for
$\zeta$ (equivalent to our Eq.(\ref{zheq})), gravitational waves (equivalent
to our Eq.(\ref{heq}) for $h$), and a test scalar field $\sigma$ (which is 
known \cite{gr74} to be similar to the equation for gravitational waves ). 
When it comes to the initial conditions, the author says that they 
are ``designed to make" 
$\zeta{\dot{\varphi}}_0/H$, $h$, and $\sigma$ behave like conventionally 
normalized free fields in the remote past. In other words, the initial 
conditions for $h$ and $\sigma$ are not ``designed to make" a potentially 
vanishing factor to enter the normalization, but the initial conditions 
for $\zeta$ are. The normalization of $\zeta$ becomes now 
propotional to $1/\sqrt \epsilon$, and the power spectrum ${\cal P}_\zeta$ 
of $\zeta$ acquires the factor $\epsilon$ in the denominator. The author 
calls this divergent power spectrum of scalar metric perturbations 
a ``classic" result.

One can imagine why this divergent formula is called ``classic". Something
repeated in the literature so many times could become ``classic" more or 
less automatically, regardless of its true value. However, Weinberg has not 
explicitly stated that the ``classic" result is a correct result. 
On the contrary, the recent paper \cite{fwein}, which deals 
with gravitational waves, makes assumptions diametrically opposite to the
prescriptions of the ``classic" result (but the paper does not
explicitly say that the ``classic" result is an incorrect result). That paper 
chooses for the analysis a model with $n_t = 0$ and `tensor/scalar 
ratio' $r=1$. This choice is in agreement with conclusions of the 
quantum theory that I advocated and 
reviewed in Sec.\ref{sec:qdenp}, but it is in conflict with what is
demanded by inflationary Eq.(\ref{tsr}) (quoted also in \cite{wein}).
Indeed, the ``classic" result for the case $n_t=0$ implies the 
non-existence of the very subject of discussion, namely, relic 
gravitational waves. Although the choice $n_t=0$, $r=1$ has been made 
\cite{fwein} purely for numerical convenience, it seems to me that the 
cautious formulation ``designed to make" \cite{weinI}, together with the 
earlier \cite{wein} and more recent \cite{fwein} treatments, testify to a 
certain evolution of views on the subject of initial conditions.   

It appears that some attempts of technical derivation of the inflationary 
$\zeta$-normalization suffer from serious inaccuracies in dealing with quantum 
operators and quantum states. Certainly, it is incorrect to think that by  
``demanding that $a^{\dagger}$ and $a$ obey the standard creation and
annihilation commutation relations we get a normalization condition for
$\zeta$". To put it in the context of a medical analogy: the rules for a 
surgical operation do not identify the patient on whom you want to operate. 
Let us discuss this point in more detail.

Suppose, being (mis)guided by various inflationary prejudices, you 
decided to write, instead of Eqs.(\ref{qp1d1}), (\ref{qp2d2}),
the following asymptotic expressions for the operators ${\bf q}$, ${\bf p}$:
\begin{equation}
\label{qp1d3}
{\bf q}= \sqrt{\frac{\hbar}{2}} \frac{a_0}{a} \frac{1}{\sqrt{\gamma}}
\left[{\bf b}e^{-in(\eta-\eta_0)} 
+ {\bf b}^{\dagger}e^{in(\eta-\eta_0)} \right],
\end{equation}
\begin{equation}
\label{qp2d3}
{\bf p}= i\sqrt{\frac{\hbar}{2}} \frac{a}{a_0} \sqrt{\gamma}
\left[-{\bf b}e^{-in(\eta-\eta_0)} + 
{\bf b}^{\dagger}e^{in(\eta-\eta_0)} \right].
\end{equation}
The commutation relationships $\left[{\bf q},{\bf p} \right] =i\hbar$ dictate
exactly the same commutation relationships for the operators 
${\bf b}^{\dagger}$, ${\bf b}$ as they did for the operators 
${\bf c}^{\dagger}$, ${\bf c}$. Namely, 
$\left[ {\bf b}, {\bf b}^{\dagger}\right]=1$.

The commutation relationships for the annihilation and creation operators 
are exactly the same, but the quantum 
state $|0_s\rangle$ annihilated by ${\bf b}$,
\[
{\bf b}|0_s \rangle =0,
\]
is totally different, it is not the ground state of the 
Hamiltonian (\ref{hamdp}). Calculation of the mean square value of the 
variable $\bar{\zeta}$ and its canonically-conjugate momentum 
gives at $\eta=\eta_0$:
\[
\langle 0_s|{\bf q}^2|0_s \rangle = \frac{\hbar}{2} \frac{1}{\gamma_0},~~~~
\langle 0_s|{\bf p}^2|0_s \rangle =\frac{\hbar}{2} \gamma_0,
\]
and the factor $\sqrt {\gamma_0}$ cancels out in the uncertainty
relation: 
\[
\Delta {\bf q} \Delta {\bf p} =\frac{\hbar}{2}.
\]

The initial $rms$ value of $\zeta$ is proportional to $1/\sqrt {\gamma_0}$
and therefore contains the `zero in the denominator' factor, but this happens 
only because the quantum state $|0_s\rangle$ is an excited (multi-quantum) 
squeezed vacuum state (for more details, see \cite{gr05} and references there). 
The choice of this state as an initial state for $\zeta$-perturbations would 
make them arbitrarily large, in the limit of $\gamma_0 \rightarrow 0$, 
right from the very beginning, i.e. from the time of imposing the initial 
conditions at $\eta = \eta_0$. The multi-quantum state $|0_s\rangle$
is not a choice of the initial state that is regarded physically 
motivated. 

At the level of classical equations, inflationists have reacted to the 
visual analogy between equations (\ref{meq}), (\ref{dpeq}), rather than, say, 
to the visual analogy between equations (\ref{heq}), (\ref{zheq}). They have
modeled initial conditions for the function $\mu_S$ on the initial 
conditions for the function $\mu_T$, instead of modeling initial conditions
for $\zeta$ on the initial conditions for the g.w. function $h$. The 
inflationary initial conditions are usually written in the form 
\[
\mu_{(T,S)} = \frac{1}{\sqrt{2k}}e^{-ik\eta}~~~~~{\rm for}~~ k|\eta| \gg 1.
\]
These initial conditions are correct for gravitational waves, but are incorrect
for density perturbations. As is seen from Eq.(\ref{zeta}), these initial 
conditions for $\mu_S$ would require the 
gauge-invariant metric perturbation $\zeta$ (as well as curvature perturbation
${\cal R}$) to be arbitrarily large, in the limit of models with 
$\gamma \rightarrow 0$, right from the very beginning, i.e. from the time of 
imposing initial conditions in the short-wavelength regime. In other words,
the inflationary initial conditions are ``designed to make" the quantity 
$\zeta$ to be divergent as $1/\sqrt{\gamma}$ from the start. These incorrect 
initial conditions are used in all inflationary calculations and 
conclusions, including the latest claims of inflationists on what the 
observations ``really" tell them about inflation.  

Finally, let us assume that the `correct' Lagrangian is given by
Eq.(\ref{Ld2new}). (Surely, you may assume correct whichever Lagrangian 
you wish, because it is taxpayers \cite{lukash} who will be paying the price 
at the end of the day.) In contrast to the Lagrangians (\ref{Lg2}), 
(\ref{Ld2}), this new Lagrangian vanishes in the most interesting limit of 
models with $\gamma \rightarrow 0$. 

For this new Lagrangian, the canonical quantization would 
require us to write (instead of Eqs.(\ref{qp1d1}), (\ref{qp2d2})): 
\begin{equation}
\label{qp1d4}
{\bf q}= \sqrt{\frac{\hbar}{2}} \frac{a_0 \sqrt{\gamma_0}}{a \sqrt{\gamma}}
\frac{1}{\sqrt{\gamma_0}}\left[{\bf d}e^{-in(\eta-\eta_0)} 
+ {\bf d}^{\dagger}e^{in(\eta-\eta_0)} \right],
\end{equation}
\begin{equation}
\label{qp2d4}
{\bf p}= i\sqrt{\frac{\hbar}{2}} \frac{a \sqrt{\gamma}}{a_0 \sqrt{\gamma_0}}
\sqrt{\gamma_0}\left[-{\bf d}e^{-in(\eta-\eta_0)} + 
{\bf d}^{\dagger}e^{in(\eta-\eta_0)} \right],
\end{equation}
where $\left[{\bf q},{\bf p} \right] =i\hbar$ and 
$\left[{\bf d}, {\bf d}^{\dagger}\right]=1$.

The ground state $|0_{new}\rangle$ of the new Hamiltonian associated with this
new Lagrangian obeys the condition
\[
{\bf d}|0_{new} \rangle =0.
\]
The mean square value of the variable $\bar{\zeta}$ in this state at 
$\eta = \eta_0$ is given by
\[
\langle 0_{new}|{\bf q}^2|0_{new} \rangle = \frac{\hbar}{2} \frac{1}{\gamma_0}.
\]
Technically speaking, one could argue that although the initial $rms$ 
value of $\zeta$ is divergent as $1/\sqrt {\gamma_0}$, the divergency
takes place for the ground state (of this new Hamiltonian), not for an
excited state (of this new Hamiltonian). But this technical subtlety 
requires a vanishing Lagrangian, and in any case it does not change much 
from the physical point of view.

Indeed, independently of technical arguments, it is important to realize
that a proposal for a divergent scalar metric power spectrum in the limit of 
${\rm  n} =1$, whatever the reasons for this proposal might be, is in 
conflict with available observations. The currently derived `best fit' 
value for ${\rm n}$ is slightly lower than 1, but the data
allow ${\rm n} =1$, even if with a smaller likelihood. This means 
that for the tested spectral indeces in the vicinity
and including ${\rm n} =1$ the data are consistent with finite and small 
amplitudes at the level of the best fit amplitude (as implied by the 
quantum theory discussed in Sec.\ref{sec:qdenp}). But there is absolutely 
nothing in the data that would suggest the need for a catastrophic growth 
of the amplitude (demanded by inflationary theory) when the data are fitted 
against spectra with indeces approaching and crossing ${\rm n} =1$. 
(Surely, the same comparison of inflationary predictions with available 
observations is addressed by inflationists as ``one of the most remarkable 
successes ... confirmed by observations".) What would we conclude about 
the assumed Lagrangian (\ref{Ld2new}) whose associated Hamiltonian leads 
to predictions contradicting observations ? We ``would conclude that
it was the wrong Lagrangian" \cite{weinbB}.

The inflationary `zero in the denominator' factor (if you decided to commit 
suicide and include it in the scalar metric power spectrum) should be 
taken at the moment of time, for a given mode $n$, when the mode begins 
its superadiabatic evolution. In general, this factor is $n$-dependent,
and hence it affects not only the overall normalization of the 
spectrum, but also its shape. It enables one to `generate' a flat, or even a
blue, power spectrum of scalar perturbations by gravitational pump fields 
which in reality can never do this. It enables one to derive all sorts of wrong 
conclusions in various subjects of study, ranging from the formation of 
primordial black holes and up to perturbations in cyclic and 
brane-world cosmologies. In particular, recent claims stating that
a given ultra-modern theory predicts an ``extremely small 
$r \lesssim 10^{-24}$~" 
or such an $r$ that ``a tensor component...is far below the detection limit 
of any future experiment", mean only that the predictions were based 
on the incorrect (inflationary) formula for scalar metric perturbations, 
with the `zero in the denominator' factor. Whatever the incorrectly derived 
$r$ may be, small or large, the use of inflationary Eq.(\ref{tsr}) in data 
analysis (what, unfortunately, is regularly being done in the CMB data 
analysis \cite{sperg,page}, \cite{pcmm}) can only spoil the extraction of 
physical information on relic gravitational waves from the data.

The self-contradictory nature of conclusions based on the
inflationary formula (\ref{tsr}) is dramatically illustrated by the claimed
derivation of limits on the amount of gravitational waves from the WMAP
data \cite{peir}. The authors use Eq.(\ref{tsr}) and explicitely quote, in 
the form of their equations (17) and (18), the power-spectrum amplitude 
$\Delta_{\cal R}^2$ of inflationary scalar perturbations and the 
`tensor-to-scalar' ratio $r$: 
\[
\Delta_{\cal R}^2 = \frac{V/M_{Pl}^4}{24 \pi^2 \epsilon}, ~~~~~~~
r= 16 \epsilon.~~~~~~~~{\rm (I)}
\]   
The conclusion of this (and many other subsequent papers) is such that 
according to the likelihood analysis of the data the value of $r$ must
be small or zero. In other words, the maximum of the likelihood function 
is either at $r=0$ or, in any case, the value of the observed $r$ is 
perfectly well ``consistent with zero". Comparing this conclusion with 
the inflationary formulas (I), quoted in the 
same paper, one has to decide whether this conclusion means that the WMAP 
data are also perfectly well consistent with arbitrarily large scalar 
amplitudes $\Delta_{\cal R}^2$, and hence with arbitrarily large 
CMB anisotropies caused by density perurbations, or that the cited 
inflationary formulas are wrong and have been rejected by observations 
(see also a discussion below, in Sec.\ref{sec:temppol}). 
The CMB data and the inflationary theory of density perturbations are 
in deep conflict with each other for long time, but inflationists and 
their followers keep claiming that they are in ``almost perfect agreement". 

It seems to me that the situation in this area of physics and cosmology 
remains unhealthy for more than 25+ years. (This is my mini-version of
`an obligation to inform the public' -- from \cite{weinb03}.) It appears 
that more than 2+ generations of researchers have been `successfully' misled 
by the ``standard inflationary results". One can only hope that the 
present generation of young researchers will be smarter and more insightful.

\section{\label{sec:det}Why Relic Gravitational Waves Should Be Detectable}

The detailed analysis in previous sections is crucial for proper 
understanding of the very status of relic gravitational waves. 
Are we undertaking difficult investigations because we want to 
find an optional ``bonus", ``smoking gun", 
``limitation" on dubious theories, or we search for something 
fundamental that `must be there', and at a measurable level ? 

Strictly speaking, it is still possible that the observed CMB anisotropies 
have nothing to do with cosmological perturbations of 
quantum-mechanical origin, that is, with the superadiabatic evolution of the 
ground state of quantized perturbations. The first worry is that, even if 
the superadiabatic (parametric) mechanism is correct by itself, the `engine' 
that drove the cosmological scale factor was unfortunate and the
pump field was too weak. In this case, a relic gravitational-wave signal (as 
well as scalar perturbations of quantum-mechanical origin) could be too small 
for discovery. I think this possibility is unlikely. First, it is difficult 
to imagine an equally unavoidable mechanism -- basic laws of quantum mechanics 
and general relativity -- for the generation of the presently existing, as 
well as the processed in the past, long-wavelength cosmological perturbations. 
Second, the inevitable quantum-mechanical squeezing and the standing-wave 
character of the generated perturbations, which, among other things, should 
have resulted in the scalar metric power spectrum oscillations (see 
Sec.\ref{sec:modf}), seems to have already revealed itself \cite{bosegr} in 
the observed CMB power spectrum oscillations. (These oscillations are often
being associated --  in my opinion, incorrectly -- with baryonic acoustic
waves and are called ``acoustic" peaks). 

A second worry is that we may be wrong in extrapolating the laws of general 
relativity and quantum mechanics to extreme conditions of the very early 
Universe. Although we apply these laws in environments that are still far away 
from any Planckian or `trans-Planckian' ambiguities, it is nevertheless an 
extremely early and unfamiliar Universe (which we want to explore). 
In principle, it is possible that something has intervened and invalidated 
the equations and rules that we have used for derivation of relic 
gravitational waves. This would also invalidate the equations and rules 
used in the derivation of density perturbations, but the generation of 
scalar perturbations requires an extra hypothesis in any case. In other 
words, even if the driving cosmological `engine' was right, the employed 
quantum theory of arising perturbations could be wrong. Hopefully, this 
complication also did not take place. 

Assuming that the dangers did not materialize, we come to the conclusion 
that we have done our job properly. The theoretical calculations 
were adequate, the normalization of $h_{rms}$ on the CMB lowest-order 
multipoles was justified, and therefore the relic gravitational wave 
background must exist, and probably at the level somewhere between the 
dotted and solid lines in Fig.\ref{gwSpectrum2}. The other side of 
the same 
argument is that if we do not detect relic gravitational waves at this 
level, something really nasty, like the above-mentioned dangers, had 
indeed happened. From not seeing relic gravitational waves, we would at 
least learn something striking about the limits of applicability of our 
main theories.

It is important to stress again that what is called here the relic 
gravitational waves is not the same thing which is sometimes called the 
inflationary gravitational waves. As the name suggests, statements about 
the inflationary gravitational waves are based on the inflation theory, as 
applied to density perturbations and gravitational waves. Although, 
conceptually, the inflationary derivation of density perturbations 
is an attempt of using the mechanism of superadiabatic 
(parametric) amplification (see Sec.\ref{sec:introduction}), the actual 
implementation of this approach has led inflationists to the divergency 
in the scalar metric spectrum (see Sec.\ref{sec:infldenp}), which has 
been converted, for the purpose of ``consistency", into a statement 
about small $r$. In particular, for the ``standard inflation" 
($\gamma \equiv \epsilon =0$), the inflationary theory predicts 
(see Eq.(\ref{tsr})) the ``smoking gun of inflation" in the form of a zero 
amount of inflationary gravitational waves ($r=0$). This is in sharp contrast 
with the superadibatic (parametric) prediction of a finite and considerable 
amount of relic gravitational waves, as shown by a dotted line 
in Fig.\ref{gwSpectrum2}. 

Some recent papers, still based on the incorrect theory with Eq.(\ref{tsr}),
start making additional artificial assumptions about the ``natural" inflationary
conditions, which amounts to postulating that $\epsilon$ should not be too 
small, but should, instead, be at the level of, say, 0.02. This kind of 
argument gives some authors ``more reason for optimism" with regard to the 
detection of primordial gravitational waves. Surely, an incorrect theory can 
lead to predictions not very much different from predictions of a correct 
theory in some narrow range of parameters that are supposed to be deduced from
observations. One should remember, however, that theories are tested not only 
by what they predict but also by what they do not predict. Any observation
consistent with the parameter value other than you postulated is against your 
theory.

Since the spectrum of relic gravitational-wave amplitudes is decreasing 
towards the higher frequencies, see Fig.\ref{gwSpectrum2}, it is 
the lowest frequencies, or better to say the longest wavelenghts, that 
provide the most of opportunities for the (indirect) detection. It is 
known for long time \cite{sw, dautc, grzel, poln} that gravitational 
waves affect the CMB temperature and polarization.
The low-order CMB multipoles ($\ell \lesssim 100$) are mostly induced by 
cosmological perturbations with wavelengths ranging from 10 times longer 
and up to 10 times shorter than the present-day Hubble radius $l_H$. And 
this is the range of scales that will be in the center of our further analysis.

\section{\label{sec:polariz}Intensity and Polarization of the CMB Radiation}

A radiation field, in our case CMB, is usually characterized by four Stokes 
parameters $(I,Q,U,V)$ \cite{Chandrasekhar1960}, \cite{LL}. $I$ is the 
total intensity of radiation (or its temperature $T$), $Q$ and $U$ describe 
the magnitude and direction of linear polarization, and $V$ is the circular 
polarization. The Stokes parameters of the radiation field arriving from a 
particular direction in the sky are functions of a point $\theta, \phi$ on 
a unit sphere centered on the observer. The metric tensor $g_{ab}$ on 
the sphere can be written as:
\begin{eqnarray}
d \sigma^{2} = g_{ab} dx^a dx^b = d \theta^{2} +\sin^{2} \theta d\phi^{2}.
\label{sphmetr}
\end{eqnarray}
The Stokes parameters are also functions of the frequency of radiation $\nu$, 
but angular dependence is more important for our present discussion.

The Stokes parameters are components of the polarization tensor
$P_{ab}$ \cite{LL} which can be written
\begin{eqnarray}
P_{ab}(\theta,\phi) = \frac{1}{2}
\left(\begin{array}{c}~~ I + Q ~~~~~~~~~~ -(U - iV)\sin{\theta} \\
-(U + iV)\sin{\theta} ~~~~~ (I - Q)\sin^{2}{\theta}
\end{array}\right).
\label{PolarizationtensorPab}
\end{eqnarray}
As every tensor, the polarization tensor $P_{ab}$ transforms under 
arbitrary coordinate transformations on the sphere, but some quantities 
remain invariant. 

We can build invariants, linear in $P_{ab}$ and its derivatives, with the 
help of the tensor $g_{ab}$ and the unit antisymmetric pseudo-tensor 
$\epsilon^{ab}$. First two invariants are easy to build:
\begin{eqnarray}
I(\theta,\phi) = g^{ab}(\theta,\phi)P_{ab}(\theta,\phi), ~~~
V(\theta,\phi) = i\epsilon^{ab}(\theta,\phi)P_{ab}(\theta,\phi).
\label{IandV}
\end{eqnarray}
Two other invariants involve second covariant derivatives of 
$P_{ab}$ \cite{bgp}:
\begin{eqnarray}
E\left(\theta,\phi\right)= -2\left(P_{ab}(\theta,\phi)\right)^{;a;b},~~~
B\left(\theta,\phi\right)=-2\left(P_{ab}(\theta,\phi)\right)^{;b;d}
\epsilon^{a}_{~d},
\label{EandB}
\end{eqnarray}

Being invariant, quantities $E$, $B$ do not mix with each other. To calculate
$E$ and $B$ in a given point $\theta, \phi$ we do not need to know the 
polarization pattern all over the sky, but we do need to know the derivatives 
of $P_{ab}$ at that point. Whatever the numerical values of $E$ or $B$ in 
a given point are, $E$ and $B$ will retain these values under arbitrary 
coordinate transformations (smoothly reducable to an ordinary rotation). 
The invariant $B$ is built with the help of a 
pseudo-tensor $\epsilon_{ab}$, and therefore $B$ is a pseudo-scalar rather 
than an ordinary scalar. While $E$ does not change sign under a coordianate
reflection, $B$ does. With $B$ one can associate the notion of 
chirality, or handedness (compare with polarization tensors 
(\ref{pctensors}), (\ref{ptensors4})). Clearly, if given cosmological 
perturbations are such that they themselves are incapable of supporting 
the handedness, it will not arise in the CMB polarization which these 
perturbations are responsible for.

The invariant quantities $I$, $V$, $E$, $B$ can be expanded over 
ordinary spherical harmonics $Y_{\ell m}(\theta,\phi)$, 
$Y_{\ell m}^* = (-1)^m Y_{\ell , -m}$: 
\begin{subequations}
\label{multipolecoefficients}
\begin{eqnarray}
I(\theta,\phi) &=& \sum_{\ell
=0}^{\infty}\sum_{m=-\ell}^{\ell}a_{\ell m}^TY_{\ell
m}(\theta,\phi),
\label{multipolecoefficientsI} \\
V(\theta,\phi) &=& \sum_{\ell =0}^{\infty}\sum_{m=-\ell }^{\ell
}a_{\ell m}^VY_{\ell m}(\theta,\phi),
\label{multipolecoefficientsV}\\
E(\theta,\phi) &=& \sum_{\ell =2}^{\infty}\sum_{m=-\ell }^{\ell }
\left[\frac{(\ell +2)!}{(\ell -2)!}\right]^{\frac{1}{2}} a_{\ell
m}^EY_{\ell m}(\theta,\phi),
\label{multipolecoefficientsE} \\
B(\theta,\phi) &=& \sum_{\ell =2}^{\infty}\sum_{m=-\ell }^{\ell }
\left[\frac{(\ell +2)!}{(\ell -2)!}\right]^{\frac{1}{2}} a_{\ell
m}^BY_{\ell m}(\theta,\phi).
\label{multipolecoefficientsB}
\end{eqnarray}
\end{subequations}
The set of (complex) multipole coefficients $(a_{\ell m}^T, a_{\ell m}^V, 
a_{\ell m}^E, a_{\ell m}^B)$ completely characterizes the intensity and
polarization of the CMB. 

The same multipole coefficients participate in the expansion of the tensor
$P_{ab}$ itself, not only in the expansion of its invariants, but the 
expansion of $P_{ab}$ requires the use of generalized spherical functions, 
the so-called spin-weighted or tensor spherical harmonics 
\cite{Zaldarriaga1997, Kamionkowski1997}. For example, the tensor 
$P_{ab}$ can be written as
\begin{eqnarray}
P_{ab} &=& \frac{1}{2}\sum_{\ell =0}^{\infty}\sum_{m=-\ell }^{\ell
}\left(g_{ab} a_{\ell m}^T
- i\epsilon_{ab} a_{\ell m}^V\right)Y_{\ell m}(\theta,\phi)\nonumber \\
&& + \frac{1}{\sqrt{2}}\sum_{\ell=2}^{\infty}\sum_{m=-l}^{l}
\left( -a_{\ell m}^E Y_{(\ell m)ab}^G(\theta,\phi) + a_{\ell
m}^BY_{(\ell m)ab}^C(\theta,\phi)\right),
\nonumber
\end{eqnarray}
where $Y_{(\ell m)ab}^G(\theta,\phi)$ and $Y_{(\ell m)ab}^C(\theta,\phi)$ 
are the `gradient' and `curl' tensor
spherical harmonics \cite{Kamionkowski1997}.

In what follows, we will not be considering the circular polarization $V$,
and we will sometimes denote the multipole coefficients collectively by 
$a_{\ell m}^X$, where $X= I, E, B$. These coefficients are to be found 
from solutions to the radiative transfer edquations in a slightly perturbed 
universe. 

\section{\label{sec:radtr}Radiative Transfer in a Perturbed Universe}

Polarization of CMB arises as a result of Thompson scattering of the initially
unpolarized light on free electrons residing in a slightly perturbed universe, 
Eq.(\ref{FRWmetric}).  Following 
\cite{Chandrasekhar1960}, \cite{Basko1980}, \cite{poln}, it is convenient
to describe Stokes parameters in terms of a 3-component
symbolic vector ${\bf \hat{n}}$:
\begin{eqnarray}
{\bf \hat{n}} = \left(\begin{array}{c} \hat{n}_1 \\ \hat{n}_2 \\
\hat{n}_3
\end{array}\right) =
\frac{c^{2}}{4 \pi \hbar \nu^{3}} \left(\begin{array}{c}
I+Q\\I-Q\\-2U
\end{array}\right).
\label{symbolicvectorstokesparameters}
\end{eqnarray} 

In the zero-order approximation, we assume that all $h_{ij}=0$ and
the CMB radiation field is fully homogeneous, isotropic, and
unpolarized. Then,
\begin{eqnarray}
{\bf \hat{n}}^{(0)} = n_0(\nu a(\eta)){\bf \hat{u}},
\label{zerothordersolution}
\end{eqnarray}
where
\begin{eqnarray}
{\bf \hat{u}} = \left(\begin{array}{c}1 \\ 1 \\0
\end{array}\right).
\nonumber
\end{eqnarray}
In the presence of metric perturbations, we write
\begin{eqnarray}
{\bf \hat{n}} = {\bf \hat{n}}^{(0)} + {\bf \hat{n}}^{(1)},
\end{eqnarray}
where ${\bf \hat{n}}^{(1)}$ is the first order correction. The functions
${\bf \hat{n}}^{(1)}$ depend on $(\eta,x^i,\tilde{\nu},e^i)$, where 
$\tilde{\nu} = \nu a(\eta)$ and $e^i$ is a unit spatial vector along the 
photon's path. Our final goal is to predict, with as much completeness as 
possible, the values of the Stokes parameters at the time of observation 
$\eta = \eta_R$.

The functions ${\bf \hat{n}}^{(1)}$ satisfy the linear version of
the radiation transfer equations. These equations can be written \cite{bgp}
\begin{eqnarray}
&&\left[ \frac{\partial }{\partial \eta} + q(\eta)+
e^i\frac{\partial}{\partial x^i} \right]
{\bf \hat{n}}^{(1)}(\eta,x^i,\tilde{\nu},e^i) = \nonumber \\
&& = ~
\frac{f(\tilde{\nu})n_0(\tilde{\nu})}{2}e^ie^j\frac{\partial
h_{ij}}{\partial \eta}{\bf \hat{u}} + q(\eta) \frac{1}{4\pi}\int
d\Omega' ~ {\bf \hat{P}}(e^i ; {e'^j}){\bf
\hat{n}}^{(1)}(\eta,x^i,\tilde{\nu},e'^j).\nonumber \\
\label{firstorderradiativetransfer}
\end{eqnarray}
The astrophysical inputs from unpolarized radiation and free 
electrons are described, respectively, by $f(\tilde{\nu})n_0(\tilde{\nu})$ and
$q(\eta)$, while the scattering process is described by 
the Chandrasekhar matrix ${\bf \hat{P}}(e^i ; {e'^j})$.
The input from the gravitational field (metric) perturbations is 
given by the combination $e^ie^j{\partial h_{ij}}/{\partial \eta}$.
Certainly, when all $h_{ij} =0$, all ${\bf \hat{n}}^{(1)}$ vanish if they 
were not present initially, which we always assume. 

The combination 
$e^ie^j{\partial h_{ij}}/{\partial \eta}$ gives rise to disparate 
angular structures for gravitational waves and density perturbations.
Let us consider a particular Fourier mode with the wavevector 
${\bf n} = (0,~0,~n)$. The polarization tensors (\ref{pctensors}) of 
gravitational waves generate the structure \cite{bgp}
\begin{eqnarray}
e^ie^j\stackrel{s}{p}_{ij}({\bf n}) = (1-\mu^{2})e^{\pm2i\phi},
\label{eep}
\end{eqnarray}
where $\mu = \cos{\theta}$ and the $\pm$ signs correspond to the left 
$s=1=L$ and right $s=2=R$ polarization states, respectively. 
Solving Eq.(\ref{firstorderradiativetransfer}) for 
${\bf \hat{n}^{(1)}}$ in terms of a series over $e^{im\phi}$, one
finds that the terms proportional to $e^{\pm2i\phi}$ are retained while 
other components $e^{im\phi}$ are not arising. This is because the 
Chandrasekhar matrix does not create any new $m\phi$ dependence.

In contrast to gravitational waves, the same metric combination with 
polarization tensors (\ref{ptensors4}) of density perturbations produces 
the structure which is $\phi$-independent. Although in the case of 
density perturbations, Eqs.(\ref{firstorderradiativetransfer}) contain one 
extra term, proportinal to the electron fluid velocity, $e^iv_i=-i\mu v_b$ 
(expressible in terms of metric perturbations via perturbed Einstein 
equations), this term is also $\phi$-independent. Since the invariant $B$ 
depends on the derivative of ${\bf \hat{n}^{(1)}}$ over $\phi$, 
one arrives at the conclusion that $B=0$ for density perturbations 
and $B \neq 0$ for gravitational waves. These results for one special 
Fourier mode ${\bf n} = (0,~0,~n)$ can then be generalized to any arbitrary 
${\bf n}$. As we anticipated on the grounds of handedness, gravitational 
waves can generate the $B$-mode of CMB polarization, but density perurbations 
can not \cite{Zaldarriaga1997,Kamionkowski1997,Hu1997b,Seljak1997a}.

\section{\label{sec:stpow}Statistics and Angular Correlation Functions}

The linear character of the radiation transfer equations 
(\ref{firstorderradiativetransfer}) makes the randomness of $h_{ij}$
being inhereted by ${\bf \hat{n}}^{(1)}$. A consistent handling 
of Eq.(\ref{firstorderradiativetransfer}) requires us to use the spatial 
Fourier expansion
\begin{eqnarray}
{\bf \hat{n}}^{(1)}(\eta,x^i,\tilde{\nu},e^i) = \frac{
\mathcal{C}}{(2\pi)^{3/2}}\int\limits_{-\infty}^{+\infty}~\frac{d^{3}{\bf
n}}{\sqrt{2n}}\sum_{s=1,2}\left[{\bf {\hat{n}}}^{(1)}_{{\bf
n},s}(\eta,\tilde{\nu},e^i)e^{i{\bf n \cdot
x}}\stackrel{s}{c}_{\bf n}+ {{\bf {\hat{n}}}^{(1)*}_{{\bf
n},s}}(\eta,\tilde{\nu},e^i)
e^{-i{\bf n \cdot x}}\stackrel{s}{c}_{\bf n}^*\right],\nonumber \\
\label{Fouriern}
\end{eqnarray}
where random coefficients $\stackrel{s}{c}_{\bf n}$ are the same 
entities which enter the metric field Eq.(\ref{fourierh}). The CMB 
anisotropies are random because the underlying metric perturbations are 
random.

In the strict quantum-mechanical version of the theory, where 
$\stackrel{s}{c}_{\bf n}$ are quantum-mechanical operators, the CMB field
${\bf \hat{n}}^{(1)}$ itself becomes a quantum-mechanical operator.
If the initial quantum state of the system is chosen to be the ground 
state $\left|0\right>$, $\stackrel{s}{c}_{\bf n}\left|0\right>=0$, all 
statistical properties of the system are determined by this choice. 

Note the extra degree of uncertainty that we will have to deal with. Usually, 
the signal is deterministic, even if totally unknown, and the randomness 
of the observed outcomes arises at the level of the measurement process, as a 
consequence of the uncontrollable noises. In our problem, if cosmological 
perturbations do indeed have quantum-mechanical origin, the signal itself 
is inherently random and is characterized by a quantum state, or a 
wave-function, or a probability distribution function. At best, we can 
predict only a probability distribution function for possible CMB maps 
(outcomes). This is true even if the dynamics and cosmological parameters 
are strictly fixed and the measurement process is strictly noiseless. 

Each mode of cosmological perturbations has started its life in the 
initial vacuum state. This state is Gaussian. In course of time it evolved 
into a squeezed vacuum state. Squeezed vacuum states retain the Gaussianity, 
even though developing the strongly unequal variances in amplitudes and 
phases. Therefore, the actually observed coefficients $a_{\ell m}^X$ 
(our own realisation of the CMB map belonging to the theoretical 
ensemble of CMB maps) are supposed to be drawn from the zero-mean Gaussian 
distributions for $a_{\ell m}^X$. If the observed CMB map looks a bit 
strange to you, due to the presence, for example, of some hints on `axes' 
and `voids', this is not necessarily an indication of non-Gaussianity of the 
underlying cosmological perturbations. Even if the observed CMB map 
consists entirely of your own images, this is not a proof 
of non-Gaussianity. It is a legitimate procedure to postulate some sort of 
a non-Gaussian distribution, by introducing a new parameter $f_{NL}$, and 
try to find $f_{NL}$ from the data of a single actually observed map. 
But it is even more a legitimate procedure to insist 
on Gaussianity of perturbations, because it follows from the very deep 
foundations explained above, and therefore regard $f_{NL} \equiv 0$ by 
definition. From this position, if the set of the observed coefficients 
$a_{\ell m}^X$ looks a bit strange to you, this is simply because our 
Universe is not as dull and `typical' as you might expect. 

In this paper, we simplify the problem and treat cosmological fields 
classically rather than quantum-mechanically. We also make mild statistical 
assumptions. Specifically, we assume that
classical random comples numbers $\stackrel{s}{c}_{\bf n}$ satisfy a 
limited set of statistical requirements:
\begin{eqnarray}
\langle \stackrel{s}{c}_{\bf n}\rangle = \langle
\stackrel{s'}{c}^*_{\bf n'}\rangle = 0, ~~~ \langle
\stackrel{s}{c}_{\bf n} \stackrel{s'}{c}^*_{\bf n'}\rangle =
\langle \stackrel{s}{c}^*_{\bf n} \stackrel{s'}{c}_{\bf n'}\rangle
= \delta_{ss'}\delta^{(3)}({\bf n} - {\bf n'}), ~~~ \langle
\stackrel{s}{c}_{\bf n} \stackrel{s'}{c}_{\bf n'}\rangle = \langle
\stackrel{s}{c}^*_{\bf n} \stackrel{s'}{c}^*_{\bf n'}\rangle = 0,
\label{statCs}
\end{eqnarray}
where the averaging is performed over some probability distributions. 
We do not even assume outright that these distributions are Gaussian. 
The rules (\ref{statCs}) are sufficient for the most of our further 
calculations.

We want to know the value of quantities $a_{\ell m}^X$  at the time
of observation $\eta = \eta_R$. To find these quantities we have to 
integrate Eq.(\ref{firstorderradiativetransfer}) over time,
with all the astrophysical and gravitational inputs taken into account.
The derived  coefficients $a_{\ell m}^X$ are random, 
because the participating coefficients $\stackrel{s}{c}_{\bf n}$ are 
random. We can calculate various correlation functions of the CMB by 
calculating the quantities 
$\langle a^{X*}_{\ell m}a^{X'}_{\ell' m'} \rangle$. Using the
rules (\ref{statCs}), one can show that these quantities take the form
\begin{eqnarray}
\langle a^{X*}_{\ell m}a^{X'}_{\ell' m'} \rangle =
C_{\ell}^{XX'}\delta_{\ell\ell'}\delta_{mm'},
\label{B5}
\end{eqnarray}
where $C_{\ell}^{XX'}$ depend on the gravitational mode functions and 
astrophysical input. $C_{\ell}^{XX'}$ are calculable as general
theoretical expressions. 

The quantities $C_{\ell}^{XX'}$ are 
called the multipoles of the corresponding CMB power spectrum $XX'$.  
As usual, the power spectrum of a field contains less information than 
the field itself, but we will mostly ignore this loss of information. Also, 
it is worth noting that practically any feature in the CMB power spectrum 
can be ``predicted" and ``explained" solely by the properly adjusted 
primordial spectrum of cosmological perturbations. But we will not go
along this line and will stick to simple power-law primordial spectra.

\begin{figure}
\begin{center}
\includegraphics[width=6cm]{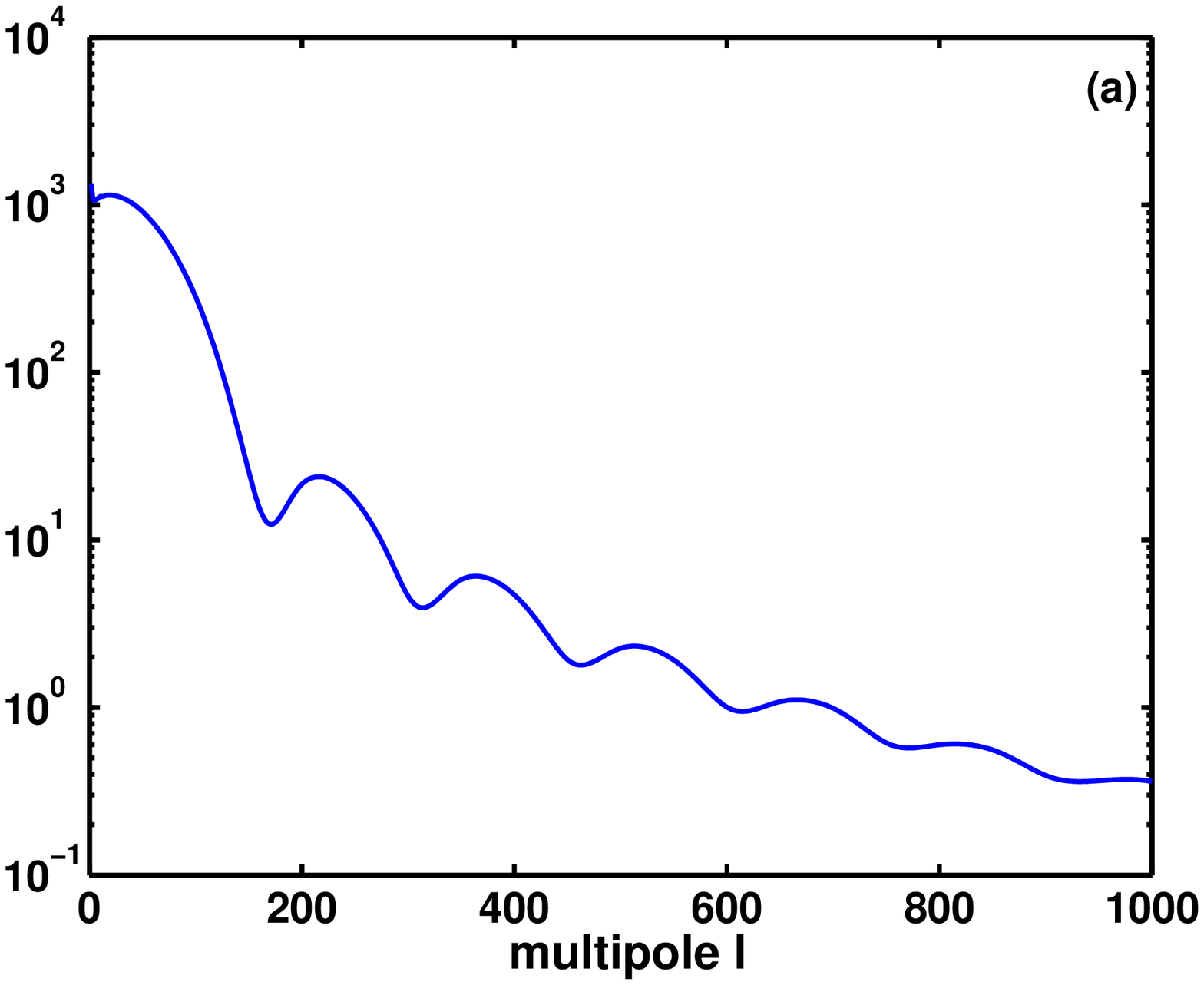}\qquad
\includegraphics[width=6cm]{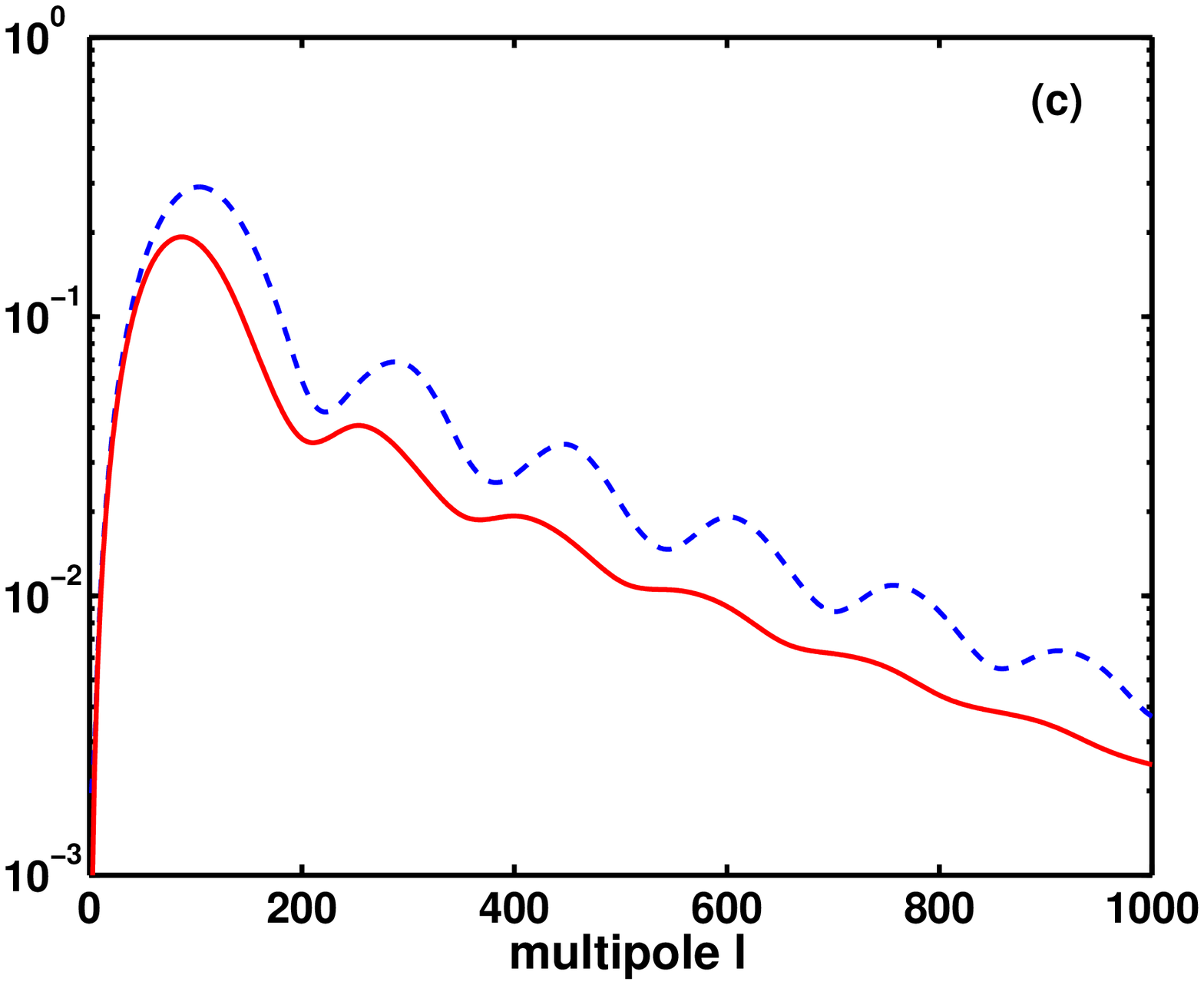}
\\
\includegraphics[width=6cm]{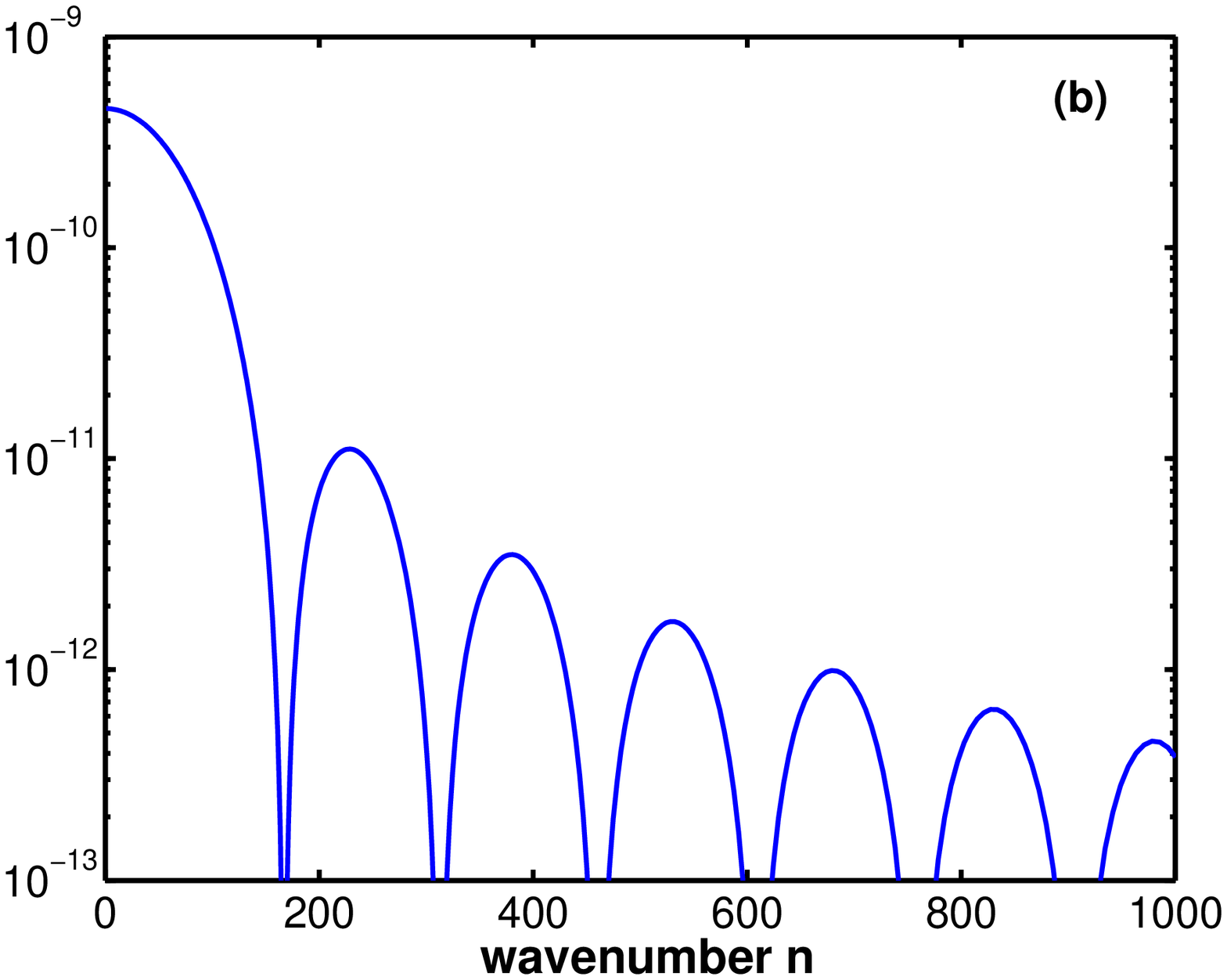}\qquad
\includegraphics[width=6cm]{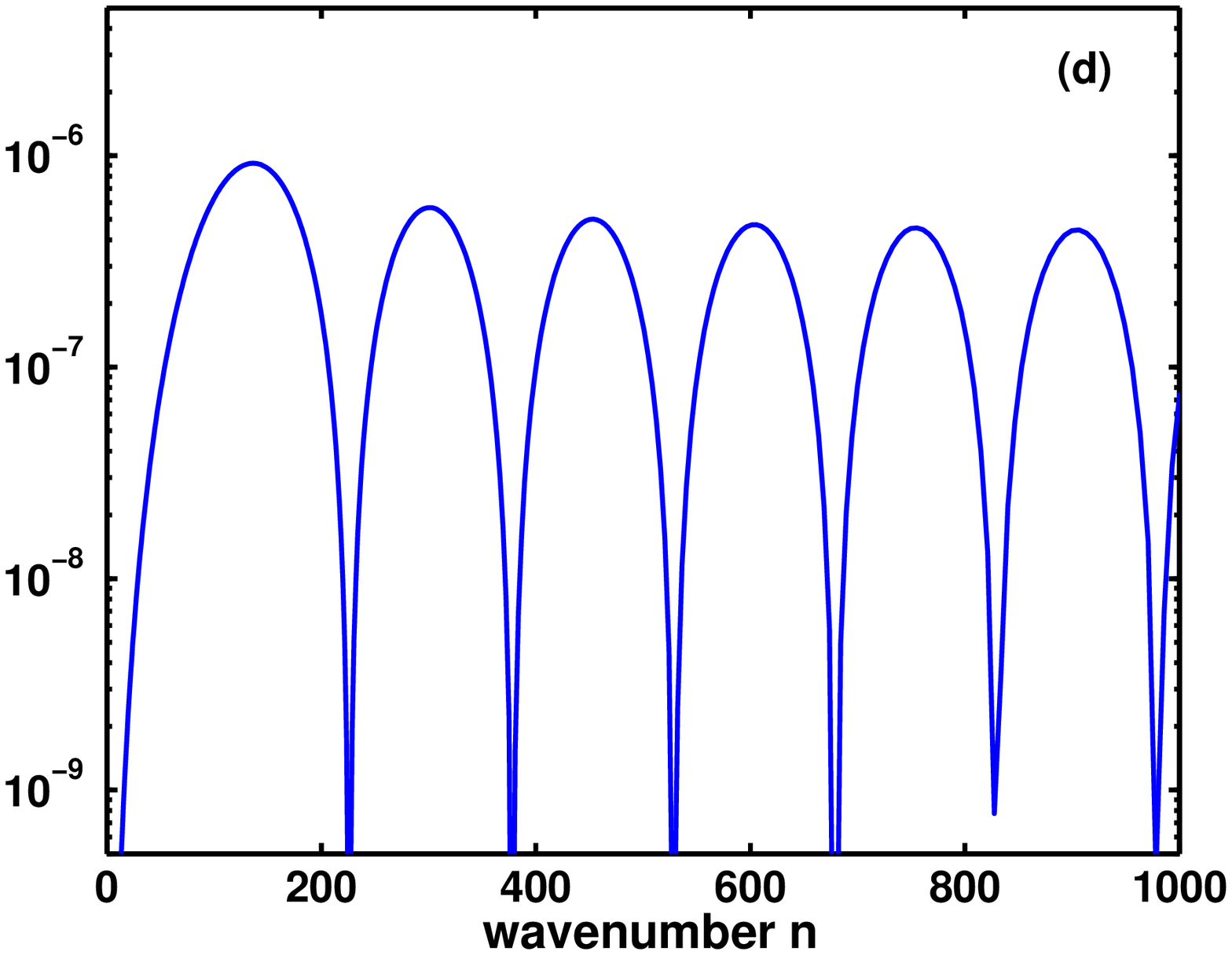}
\end{center}
\caption{The left panel shows (a) the power spectrum of
temperature anisotropies $ \ell (\ell+1)C_{\ell}^{TT}$ (in $~\mu
\textrm{K} ^{2}$) generated by (b) the power spectrum of g.w.\
metric perturbations $hh$ (\ref{gwpower}), $\beta =-2$. The right panel
shows (c) the power spectra of polarization anisotropies $\ell
(\ell+1)C_{\ell}^{BB}$ (solid line) and $\ell
(\ell+1)C_{\ell}^{EE}$ (dashed line), panel (d) shows the power
spectrum of the first time derivative of the same g.w.\ field,
$h'h'$.} 
\label{figure_recombination4}
\end{figure}

\begin{figure}
\begin{center}
\includegraphics[width=6cm]{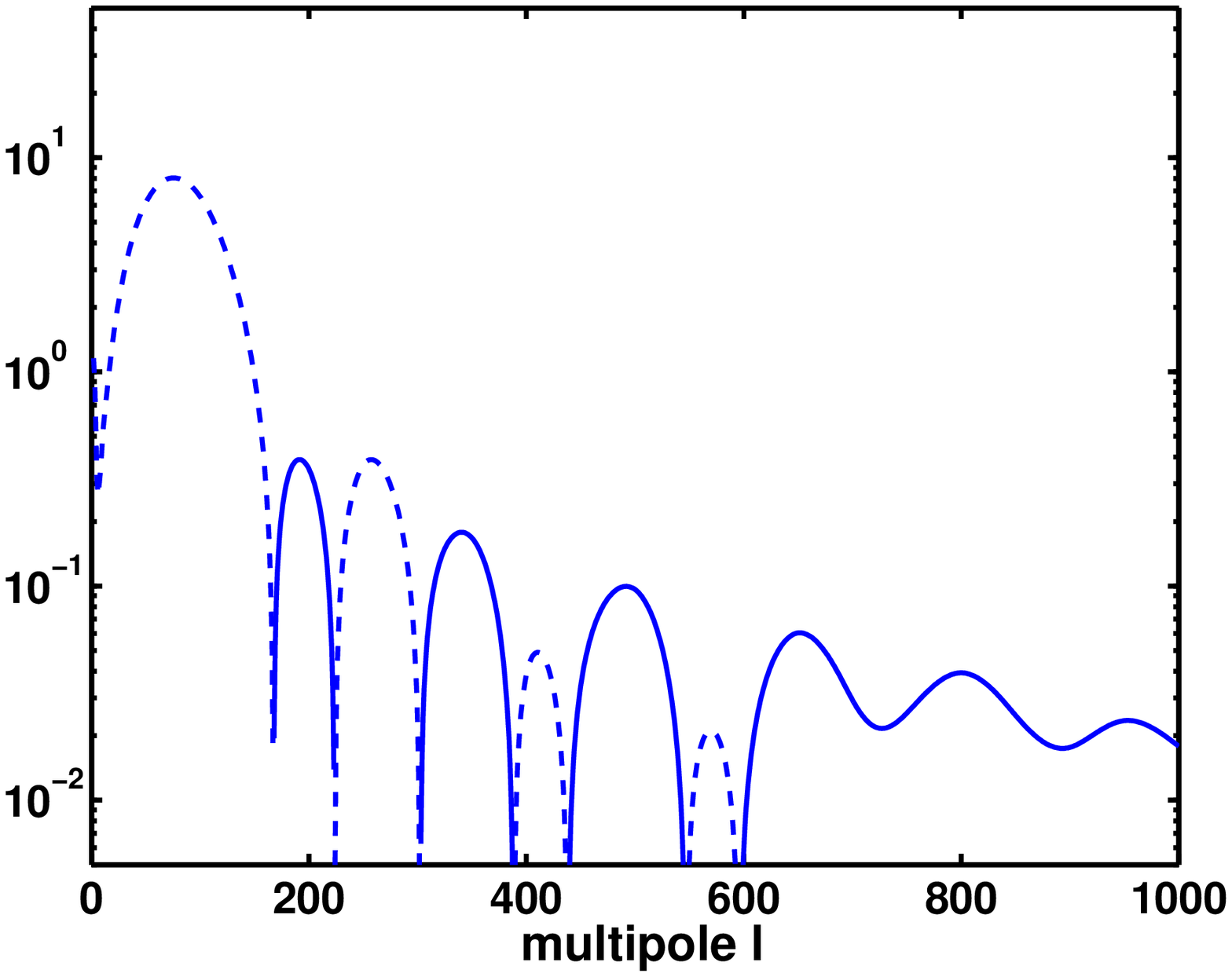}
\\
\includegraphics[width=6cm]{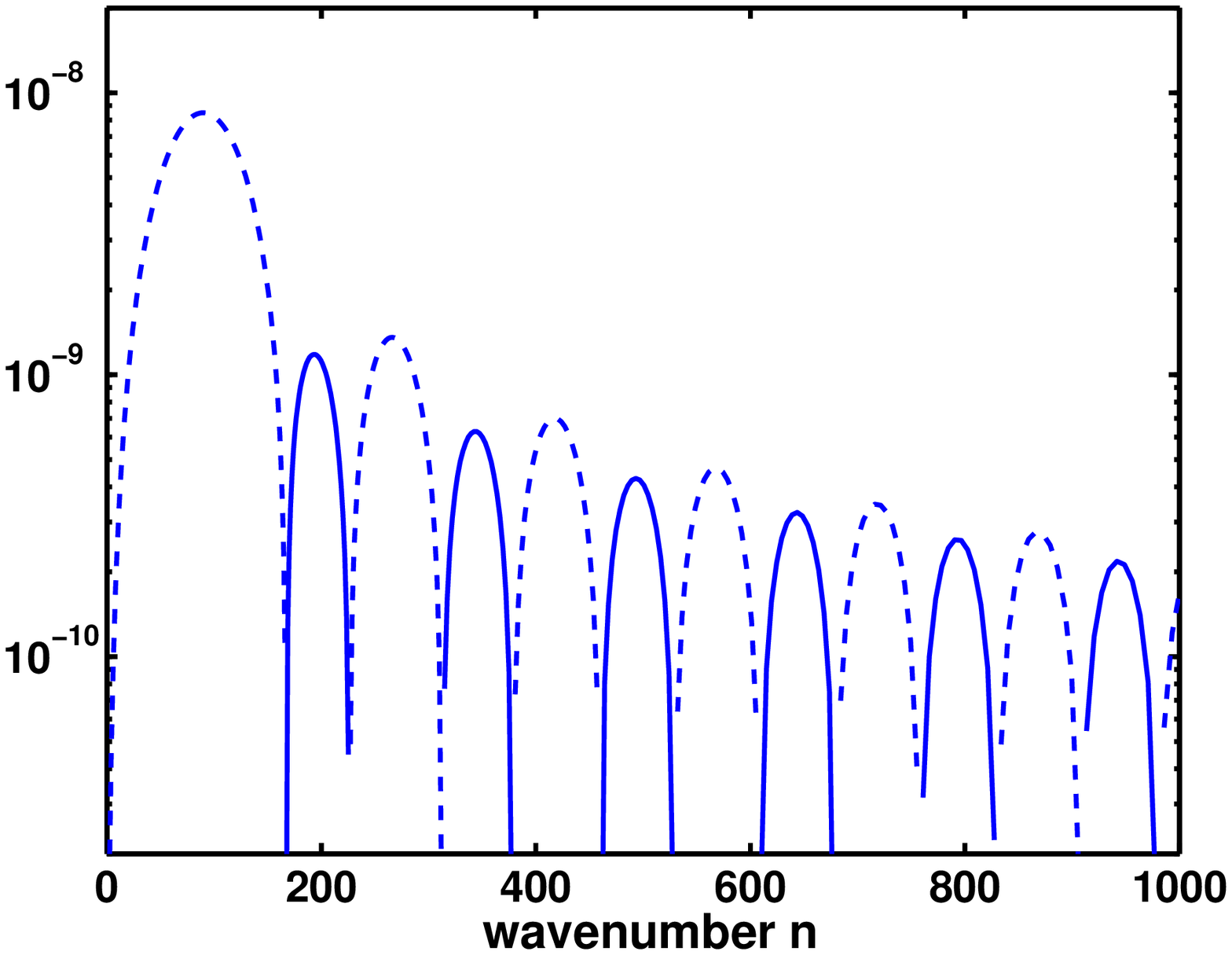}
\end{center}
\caption{The bottom panel shows the cross-power spectrum $hh'$ of
gravitational waves, whereas the top panel shows the angular power
spectrum $\ell(\ell+1) C^{TE}_{\ell}$ caused by these waves. The
negative values of these functions are depicted by broken lines.}
\label{powcorrTE}
\end{figure}

\begin{figure}
\begin{center}
\includegraphics[width=10cm]{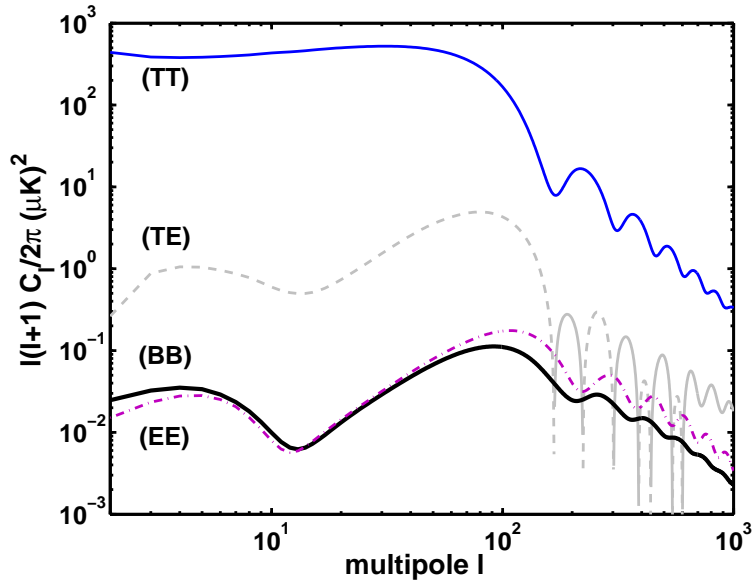}
\end{center}
\caption{The summary of CMB temperature and polarization
anisotropies due to relic gravitational waves with $n=1.2$ and
$R=1$.} 
\label{figureCMBgw}
\end{figure}

In the case of relic gravitational waves, the general shape of today's 
CMB power spectra and their features in the $\ell$-space are almost in 
one-to-one correspondence with the processed metric power spectra and their 
features in the wavenumber $n$-space. The processed metric power spectra 
should be taken at the time of decoupling of CMB. The $TT$ power spectrum is 
determined by the power spectrum of the metric itself, $hh$. 
The $EE$ and $BB$ power spectra are largely determined by 
the power spectrum of the metric's first time-derivative, $h'h'$. And $TE$ 
power spectrum is determined by the cross power spectrum of the metric and 
its first time-derivative, $hh'$. For the gravitational-wave background
with parameters indicated by dotted line in Fig.\ref{gwSpectrum2}, 
the corresponding metric and CMB power spectra are shown in 
Fig.\ref{figure_recombination4} and Fig.\ref{powcorrTE}. For the 
gravitational-wave background with parameters indicated by a solid line in 
Fig.\ref{gwSpectrum2}, the summary of CMB spectra is shown in 
Fig.\ref{figureCMBgw}. The summary 
also includes the reionization `bump' at $\ell \lesssim 12$. (More details 
about these figures can be found in \cite{bgp}).

All CMB power spectra have been calculated as averages over a theoretical 
ensemble of all possible realizations of the CMB field. In its turn, this 
randomness of CMB anisotropies ensues from the randomness of the 
gravitational field coefficients 
$\stackrel{s}{c}_{\bf n}$, $\stackrel{s}{c}_{\bf n}^{*}$. 
The characteristic parameters of a stochastic field, such as 
its mean values and variances, are, by definition, averages over the 
ensemble of realizations. However, in CMB observations, we have access to 
only one realization of this ensemble, which can be thought of as a 
single observed set of coefficients $a_{\ell m}^X$. Is it possible to find the
parameters of a stochastic process by studying only one realization of this 
process ? 

The answer is yes, if the correlations of the stochastic process 
decay sufficiently quickly at large separations in time or space where the 
process is defined \cite{yaglom}. The process allowing the derivation of its
true parameters, with probability arbitrarily close to 1, from a single 
realization is called ergodic. For example, the distribution of 
galaxies, or a stochastic density field, in an infinite 3~-~space 
(our Universe) 
may be ergodic, and then by studying a single realization of this 
distribution we could extract the true parameters of the underlying random 
process. In the theory that we are discussing here, the randomness
of the linear density field is also described by the random coefficients 
$\stackrel{s}{c}_{\bf n}$, $\stackrel{s}{c}_{\bf n}^{*}$ appearing in the 
metric perturbations.

If the process is non-ergodic, there will be an inevitable uncertainty
surrounding the parameter's estimation derived from a single realization. 
This uncertainty should not be mis-taken for the `cosmic variance', often 
quoted and plotted on the observational graphs. The cosmic variance is a 
mathematically correct statement about the size of the variance of 
the $\chi^2$-distribution for $2\ell+1$ independent Gaussian variables. 
There is nothing particularly cosmic in the cosmic variance. The 
elementary theorem, called cosmic variance, has nothing to say 
about the ergodicity or non-ergodicity of a given stochastic process. 
In particular, the universal validity of cosmic variance cannot prevent the 
extraction of exact parameters from the observation of a single realization 
of the stochastic density field in our Universe, if the density field is 
ergodic, -- and it might be ergodic.

The problem is, however, that on a compact 2-sphere, where the random CMB is 
defined, ergodic processes do not exist. We will always be facing some 
uncertaintly related to non-ergodicity. The size of this uncertainty about
the derived parameter depends on the statistics and employed estimator. 
Under some conditions, this uncertainty is close, numerically, to the size 
of the usually quoted cosmic variance \cite{grmar}. This discussion is 
important, because we are now approching the observational predictions and 
the ways of discovering relic gravitational waves in the CMB data.

\section{\label{sec:temppol}Temperature-Polarization Cross-Correlation Function}

The numerical levels of primordial spectra for tensor and scalar metric 
perturbations are approximately equal. This means that the amplitudes of 
those Fourier modes $n$ which have not started yet their short-wavelength 
evolution are numerically comparable, and the metric mode functions are 
practically constant in time. In contrast, in the short-wavelength 
regime, the amplitudes of gravitational waves adiabatically decrease, 
while the amplitudes of gravitational field perturbations associated 
with density parturbations may grow. Specifically, at the 
matter-dominated stage, the function $h_l(\eta)$ grows and overtakes 
$h(\eta)$, which remains constant. 

In the context of CMB anisotropies, the crucial time is the epoch of 
decoupling. The wavelenghts of modes with $n \lesssim 100$ were comfortably 
longer than the Hubble radius at the decoupling. The influence of these 
modes, both in gravitational waves (g.w.) and density perturbations (d.p.), 
have been projected into today's $XX'$ anisotropies at $\ell \lesssim 100$. 
In this interval of $\ell$, the g.w. contribution is not a 
small effect in comparison with the d.p. contribution. The g.w. contribution 
to the CMB power spectra is illustrated in Fig.\ref{figure_recombination4}, 
Fig.\ref{powcorrTE}, and Fig.\ref{figureCMBgw}.

One way of detecting relic gravitational waves is based on measuring the 
$BB$ auto-correlation. This method is clean, in the sense that density 
perturbations do not intervene, but the expected signal is very
weak, see Fig.\ref{figure_recombination4}, Fig.\ref{figureCMBgw}. 

We propose \cite{bgp} to concentrate on the $TE$ cross-correlation 
(without, of course, neglecting the $BB$ searches). The $TE$ signal 
is about two orders of magnitude stronger than 
the $BB$ signal, and the use of a cross-correlation is always better 
than an auto-correlation, in the sense of fighting the noises. The special 
feature allowing to distinguish the g.w. part of $TE$ from 
the d.p. part of $TE$ is the difference in their signs. In the
interval $\ell \lesssim 100$ the $(TE)_{gw}$ must be negative
(see Fig.\ref{powcorrTE}),
while the $(TE)_{dp}$ must be positive at lowest $\ell's$ and up to, at least, 
$\ell \approx 50$. This difference in sign of $TE$ correlation functions is 
the consequence of the difference in sign of the g.w. and d.p. metric 
cross-power spectra $hh'$ in the interval $n \lesssim 70$ \cite{bgp}.   
(The difference in sign of $QT$ correlation functions is discussed in
the earlier paper \cite{cct}.)

\begin{figure}
\begin{center}
\includegraphics[width=8cm]{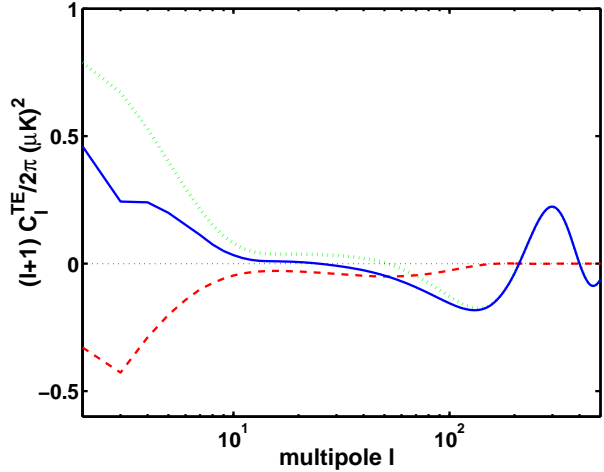}
\end{center}
\caption{The dotted line shows the contribution of density
perturbations alone, while the dashed line shows the contribution of
gravitational waves alone. The solid line is the sum of these
contributions. It is seen from the graph that the inclusion of
g.w.\ makes the total curve to be below the d.p.\ curve.}
\label{te_just_adding_gw}
\end{figure}

An example of expected g.w. and d.p. contributions to the $TE$ correlation 
function is shown in Fig.\ref{te_just_adding_gw}. To include reionization 
and enhance the 
lowest $\ell's$, we plot the function $(\ell+1)/2\pi~C_{\ell}^{TE}$ rather 
than the usual $\ell (\ell+1)/2\pi~C_{\ell}^{TE}$. The g.w. and d.p. metric 
power spectra are normalized in such a way that they give $R=1$, where 
\[
R \equiv \frac{C_{\ell=2}^{TT}(gw)}{C_{\ell=2}^{TT}(dp)}.
\]
More specifically, they give equal contributions to the total `best fit' 
\cite{page} temperature quadrupole $C_{\ell=2}^{TT}$. The dashed line in
Fig.\ref{te_just_adding_gw} shows the effect of the g.w. background marked by 
dotted line in Fig.\ref{gwSpectrum2}. The dotted line in 
Fig.\ref{te_just_adding_gw} shows the effect 
of d.p. metric perturbations, with the same primordial spectral index, 
${\rm n} =1$. The sum of the two contributions is shown by a solid line. 
The fact that the two contributions may almost cancel each other does not
mean that the g.w. signature is weak and hard to measure. The g.w. signal
is in the strong deviation of the total $TE$ spectrum from the expectation
of the d.p. model. 

It is important to remember that the widely publicized ``tight 
limits" on the `tensor-to-scalar ratio' $r$ which allegedly rule out $R=1$
and any $R \neq 0$, unless $R$ is small, 
were derived by making use of the inflationary ``consistency relation" (see
Sec.\ref{sec:infldenp}). In these derivations, the g.w. spectral index 
$n_t$ is being taken from the relationship $n_t= - r/8$, which automatically 
sends $r$ to zero when $n_t$ approaches zero. If this relationship is 
regarded as an artificial extra condition on g.w. parameters, 
then the results of such a data analysis may be of some value to 
those who are interested in this {\it ad hoc} condition, but not to those
who are interested in determination of the true amount of relic 
gravitational waves. If, on the other hand, this relationship is 
regarded as part of inflationary theory, then such a data analysis is 
deeply self-contradictory. Indeed, the invariably derived conclusion, 
according to which the maximum of the likelihood function for $r$ is at 
$r=0$, or at least the value of the observed $r$ is ``consistent with zero", 
means that the most likely values of the inflationary density perturbation 
amplitudes, together 
with the CMB anisotropies induced by them, are infinitely large, or at least 
the data are ``consistent" with such an infinity  (see end of 
Sec.\ref{sec:infldenp}). It goes without saying that we are not using 
this relationship. But we do use the relationship $n_t = n_{s} -1$ which is 
a consequence of the superadiabatic generating mechanism, 
see Eq.(\ref{spectins}).

The WMAP community seems to be satisfied with the data analysis which states 
that the CMB data can be described by a small number of parameters 
which include density perturbations, but with no necessity for gravitational 
waves. The usual logic in these derivations is first to find the best-fit 
parameters assuming that only the density perturbations are present, and then 
to claim that there is no much room left for inclusion of gravitational waves. 
This looks like being satisfied with the statement that most 
of what is known about the human race can be described by a small 
number of parameters which includes one leg of individuals, but with no 
need for another. And when you propose to the data analysts that 
it is better to treat the data under the assumption that humans have two 
legs, they reply that this would be one extra parameter, and the 
proposer should be penalized for that. Anyway, in our analysis, we 
include the (inevitable) gravitational waves from the very beginning.

The total $TE$ signal is the sum of g.w. and d.p. contributions. Even if 
this sum is positive in the interval $\ell \lesssim 70$, the effect of 
gravitational waves can (and expected to) be considerable. In this case, 
the amount of gravitational waves can be estimated through the analysis of 
all correlation functions together. However, if the total $TE$ signal  
is negative in this interval of $\ell$, there is little doubt that a 
significant g.w. component is present and is responsible for the negative
signal, because the mean value of the $TE$ signal cannot be negative 
without gravitational waves (the issues of statistics are discussed below).

It is intriguing that the WMAP team \cite{page} explicitly emphasizes the 
detection of a negative correlation (i.e. anticorrelation) at the 
multipoles near $\ell \approx 30$: ``The detection of the $TE$ 
anticorrelation near $\ell \approx 30$ is a fundamental measurement 
of the physics of the formation of cosmological perturbations...". The 
motivation for this statement is the continuing concern of CMB 
observers about the so-called `defect' models of structure formation. 
These `causal' models cannot produce any correlations, positive or 
negative, at $\ell \lesssim 100$. So, a detected correlation 
near $\ell \approx 30$ is an evidence againt them. At the same time, 
the better detected $TE$ anticorrelation in the region of higher 
$\ell \approx 150$ could still be accomodated by the causal models,
and therefore this more visible feature is not a direct argument against 
these models. However, the motivation for the WMAP team's statement 
is not essential for the present discussion. Even if the available data are 
not sufficient to conclude with confidence that the excessive $TE$ 
anticorrelation at lower $\ell$'s has been actually detected, it seems that the 
WMAP's published data (together with the published statement, quoted above) 
can serve at least as an indication that this is likely to be true. In the 
framework of the theory that we are discussing here, such an $TE$ 
anticorrelation is a natural and expected feature due to relic gravitational 
waves, but of course this needs to be thoroughly investigated.  

The ensemble-averaged correlation functions do not answer all the questions.
It is necessary to know what one would get with individual realizations 
of $a_{\ell m}^X$ caused by individual realizations of random coefficients
$\stackrel{s}{c}_{\bf n}$, $\stackrel{s}{c}_{\bf n}^{*}$. The general
unbiased estimator $D_{\ell}^{XX'}$ of $C_{\ell}^{XX'}$ is given by
\begin{equation}
\label{Cest}
D_{\ell}^{XX'}= \frac{1}{2(2\ell +1)} \sum_{m=-\ell}^{\ell}
\left(a_{\ell m}^{X}a_{\ell m}^{X' *}+a_{\ell m}^{X *}a_{\ell m}^{X'}\right),
\end{equation}
where $a_{\ell m}^{X}$ depend in a complicated, but calculable, manner on
mode functions, astrophysical input, and, in general, on all random 
coefficients $\stackrel{s}{c}_{\bf n}$, $\stackrel{s}{c}_{\bf n}^{*}$.
(It was explicitely shown \cite{grmar} that the estimator $D_{\ell}^{TT}$ is not
only unbiased, but also the best, i.e. the minimum-variance, estimator among 
quadratic estimators.) One can write, symbolically,
\begin{eqnarray}
a^{X}_{\ell m} =  \int \limits_{-\infty}^{+\infty} d^{3}{\bf n}
\sum_{s=1,2}
\left[f^X_{{\ell}m}(n,s)\stackrel{s}{c}_{\bf n} +
f^{X*}_{{\ell}m}(n,s)\stackrel{s}{c}_{\bf n}^*\right].
\label{mXcomp}
\end{eqnarray}

Some features of the averaged $XX'$ functions remain the same in any 
realization, i.e. for any choice of random coefficients 
$\stackrel{s}{c}_{\bf n}$, $\stackrel{s}{c}_{\bf n}^{*}$. One example 
is the absence of $BB$ correlations for density perturbations. In the 
case of density perturbations, the $BB$ correlation function vanishes 
not just on average, but in every realization. This happens because all the 
functions $f^B_{{\ell} m}(n,s)$ in Eq.(\ref{mXcomp}) are zeros. Another 
example is the retention, in every realization, of positive sign of 
the auto-correlation functions $XX'$, with $X'=X$. Indeed, 
expression (\ref{mXcomp}) may be arbitrarily complicated, but
the estimator (\ref{Cest}) for $X'=X$ is always positive, as it is the sum of 
strictly positive terms. In different realizations the estimates will be
different, but they will always be positive. That is, in any realization,
including the actually observed one, the 
quantity $D_{\ell}^{XX}$ has the same sign as the quantity $C_{\ell}^{XX}$.  

The important question is how often the $TE$ estimations (\ref{Cest}) retain 
the same sign as the already calculated ensemble-averaged quantity 
$C_{\ell}^{TE}$. 
In other words, suppose a negative $TE$ is actually observed in the
interval $\ell \lesssim 50$. Can it be a statistical fluke of density
perturbations alone, rather than a signature of gravitational waves ? 
In general, the answer is `yes', but we will present arguments why in the 
problem under discussion the answer may be `no', or `very likely no'.

To quantify the situation, it is instructive to start from two 
zero-mean Gaussian variables, say, $a^T$ and $a^E$:
\begin{equation}
\label{2ex}
\langle a^T a^T \rangle = \sigma^2, ~~~ \langle a^E a^E \rangle = \sigma^2,~~~
\langle a^T a^E \rangle = \rho \sigma^2, ~~0 \leq \rho \leq 1.
\end{equation}
The probability density function for the product variable $a^Ta^E$ involves 
the modified Bessel function $K_0$ and is known as an exact 
expression \cite{gr96}. In the situation, like ours, where $\rho$ is 
close to 1, the probability of finding negative products $a^Ta^E$ is small. 
A crude evaluation shows that, with probabilty 68\% and higher, the 
values of $a^Ta^E$ lie in the interval $(0.3 - 1.7)$ around the positive 
mean value $\langle a^Ta^E \rangle$, whereas for $\rho=1$ the probability
of finding a negative $\langle a^Ta^E \rangle$ is strictly zero. 
For $2\ell +1$ degrees of freedom the scatter around the mean value is 
expected to be much narrower. Nevertheless, if $\rho \neq 1$, infrequent 
realizations with negative values of the product $a^Ta^E$ are still possible.

We are probably quite fortunate in our concrete case of $TE$ 
cross-correlation functions. It is true that $a_{\ell m}^T$ and 
$a_{\ell m}^E$ contain a large number of independent random coefficients 
$\stackrel{s}{c}_{\bf n}$, multiplied by different deterministic 
functions $f^T_{{\ell}m}(n,s)$, $f^E_{{\ell}m}(n,s)$. However, in practice, 
these deterministic functions are such that the integrands in 
Eq.(\ref{mXcomp}) are quite sharply peaked at $n \approx \ell$. Ideally, 
for every $\ell$, the random variables $a_{\ell m}^T$, 
$a_{\ell m}^E$ become proportional to one and the same linear combination
of random coefficients with $|{\bf n}|=n \approx \ell$. The square 
of this combination is always positive, so the sign of the estimator 
$D_{\ell}^{TE}$ would be the same as the sign of the mean value 
$C_{\ell}^{TE}$. This sign is determined by the known deterministic 
functions, not by statistics. Although this practical retention of the 
sign is not a strictly proven theorem, I think it is a very plausible 
conjecture.

If the above-mentioned conjecture is correct, the negative sign of any         
observed $TE$ correlation in the interval $\ell \lesssim 50$ is
unlikely to be a statistical fluke of density perturbations alone, it should 
be a signature of presence of gravitational waves. One especially interesting 
interval of $TE$ searches is the interval between $\ell \approx 30$ and 
$\ell \approx 50$. In this interval of $\ell$, the total signal 
$C_{\ell}^{TE}$ should be negative, whereas the contribution from density 
perturbations is still positive (see Fig.\ref{te_just_adding_gw} for a 
realistic example). The total $TE$-signal is a factor 50, or so, larger than 
the expected $BB$-signal. All the logic described above suggests that if a 
negative $TE$ is detected in this interval of $\ell$'s it must have occured 
due to relic gravitational waves.
 
\section{\label{sec:pros}Prospects of the Current and Forthcoming Observations}

It is difficult to predict the future, but it seems to me that the
Planck mission \cite{Planck}, as well as the ground-based experiments,
such as BICEP \cite{bicep1, bicep2}, Clover \cite{clover}, QUIET \cite{QUIET},  
have a very 
good chance of detecting relic gravitational waves through the $TE$ 
anticorrelation described above. A level of the expected $BB$ correlation 
is shown in Fig.\ref{figureCMBgw}. This level is natural for the discussed 
theory, but may be a little bit optimistic, as it is shown for
a relatively high primordial spectral index ${\rm n} = 1.2$. 
Observations with the help of space and ground facilities, and the measurement 
of all relevant correlation functions, should bring positive results given the
expected sensitivities of those observations. It seems 
to me that the detection of relic gravitational waves in CMB is the matter of 
a few coming years, rather than a few decades. The analysis of the latest WMAP 
data provides serious indications of the presence of relic gravitational waves 
in the CMB anisotropies, see \cite{zbg1}, \cite{zbg2}.

\section*{Acknowledgments}

I am greatful to D. Baskaran and A. G. Polnarev for collaboration on the joint 
paper \cite{bgp} that was extensively quoted in this presentation, and to 
S. Weinberg for helpful comments on this manuscript.

%

\begin{thebibliography}{10}

\bibitem{MTW}
C. W. Misner, K. S. Thorne, and J. A. Wheeler.
\newblock {\em Gravitation}, Freeman and Co., San Francisco, 1973

\bibitem{gr74}
L.~P. {Grishchuk}.
\newblock {\em Zh. Eksp. Teor. Fiz.} {\bf 67}, 825 (1974) [Sov. Phys. 
JETP {\bf 40}, 409 (1975)]; {\em Ann. NY Acad. Sci.} {\bf 302}, 439 
(1977); {\em Pis'ma Zh. Eksp. Teor. Fiz.} {\bf 23}, 326 (1976) [JETP Lett. 
{\bf 23}, 293 (1976)]; {\em Uspekhi Fiz. Nauk} {\bf 121}, 629 (1977)
[Sov. Phys. Usp. {\bf 20}, 319 {1977}]

\bibitem{sakh65} A. D. Sakharov. 
\newblock {\em Zh. Eksp. Teor. Fiz.} {\bf 49}, 345 (1965) [Sov. Phys. JETP 
{\bf 22}, 241 (1966)] 

\bibitem{schr} E. Schrodinger.
\newblock {\em Physica} {\bf 6}, 899 (1939)

\bibitem{gr93prd}
L. P. Grishchuk.
\newblock {\em Phys. Rev. D}{\bf 48}, 5581 (1993)

\bibitem{parker}
L. Parker.
\newblock {\em Phys. Rev. Lett.} {\bf 21}, 562 (1968); {\em Phys. Rev.}
{\bf 183}, 1057 (1969)

\bibitem{byz}
B. Ya. Zeldovich.
{\em Impedance and parametric excitation of coupled oscillators}, to be
published in {\em Physics-Uspekhi} (2008)

\bibitem{LLmech}
L.~D. {Landau} and E.~M. {Lifshitz}.
\newblock {\em Mechanics}. Oxford, Pergamon Press, 1975

\bibitem{zeld}
P. Paradoksov.
\newblock {\em Uspekhi Fiz. Nauk} {\bf 89}, 707 (1966) [Sov. Phys. Usp. 
{\bf 9}, 618 (1967)]

\bibitem{LL}
L.~D. {Landau} and E.~M. {Lifshitz}.
\newblock {\em {The Classical Theory of Fields}}. Oxford, Pergamon Press, 1975

\bibitem{gr93}
L. P. Grishchuk. 
\newblock {\em Class. Quant. Gravity} {\bf 10}, 2449 (1993)

\bibitem{lectnotes}
L. P. Grishchuk.
\newblock {\em Lecture Notes in Physics}, Vol.~562: Gyros, Clocks, 
Interferometers: Testing Relativistic Gravity in Space, 167 (2001)

\bibitem{grsol}
L. P. Grishchuk and M. Solokhin.
\newblock {\em Phys. Rev. D}{\bf 43}, 2566 (1991)

\bibitem{weinbB}
S. Weinberg.
\newblock{\em The Quantum Theory of Fields.} Cambridge U. Press, 1995

\bibitem{gr05}
L.~P. {Grishchuk}.
\newblock {\em Uspekhi Fiz. Nauk} {\bf 175(12)} 1289-1303 (2005) 
[Physics-Uspekhi, {\bf 48}(12) 1235-1247 (2005], arXiv:gr-qc/0504018

\bibitem{grpoln}
L. P. Grishchuk and A. G. Polnarev.
\newblock In: {\em General Relativity and Gravitation, 100 Years After the
Birth of A. Einstein}, ed. A. Held (Plenum Press, NY, 1980) Vol. 2, p. 393

\bibitem{bgp}
D. Baskaran, L. P. Grishchuk, and A. G. Polnarev.
\newblock {\em Phys. Rev. D}{\bf 74}, 083008 (2006)

\bibitem{zelnov}
Ya. B. Zeldovich and I. D. Novikov.
\newblock{\em The Structure and Evolution of the Universe.} Chicago U. Press,
1983

\bibitem{gr94}
L.~P. Grishchuk.
\newblock {\em Phys. Rev. D}{\bf 50}, 7154 (1994)

\bibitem{Lie}
L. P. Eisenhart.
{\em Continuous Groups of Transformations}. Dover, 1933

\bibitem{bard}
J. M. Bardeen.
\newblock{\em Phys. Rev. D}{\bf 22}, 1882 (1980)

\bibitem{BST}
J. M. Bardeen, P. J. Steinhardt, and M. S. Turner. 
\newblock {\em Phys. Rev. D}{\bf 28}, 679 (1983)

\bibitem{luk} V. N. Lukash. 
\newblock {\em Pis'ma Zh. Eksp. Teor. Fiz.} {\bf 31}, 631 (1980)
[JETP Lett. {\bf 6}, 596 (1980)]

\bibitem{chm} G. Chibisov and V. Mukhanov. 
\newblock {\em Mon. Not. R. Astr. Soc.} {\bf 200}, 535 (1982)

\bibitem{sas} M. Sasaki. 
\newblock {\em Prog. Theor. Phys.} {\bf 76}, 1036 (1986) 

\bibitem{unruh}
W. Unruh.
{\em Cosmological long wavelength perturbations}, arXiv:astro-ph/9802323

\bibitem{weinI}
S. Weinberg.
{\em Phys. Rev. D}{\bf 72}, 043514 (2005)

\bibitem{fwein}
R. Flauger and S. Weinberg.
{\em Phys. Rev. D}{\bf 75}, 123505 (2007) (arXiv:astro-ph/0703179)

\bibitem{wein}
S. Weinberg.
{\em Phys. Rev. D}{\bf 69}, 023503 (2004)

\bibitem{lukash}
V. N. Lukash.
{\em On the relation between tensor and scalar perturbation modes in Friedmann
cosmology,} arXiv:astro-ph/0610312

\bibitem{sperg}
D. N. Spergel {\it et al.} {\em WMAP Collaboration},
\newblock {\em Astrophys. J. Suppl.} {\bf 170}, 377 (2007)
(arXiv:astro-ph/0603449)

\bibitem{page}
L. Page {\it et al.} {\em WMAP Collaboration},
\newblock {\em Astrophys. J. Suppl.} {\bf 170}, 335 (2007)
(arXiv:astro-ph0603450)

\bibitem{pcmm}
L. Pagano, A. Cooray, A. Melchiori, and M. Kamionkowski.
\newblock {\em Red Density Perturbations and Inflationary Gravitational Waves},
arXiv:0707.2560 (In this fresh paper one can also find some previous history
of derivation and application of the inflationary `consistency relation'.) 

\bibitem{peir}
H. V. Peiris {\it et al.}
\newblock {\em Astrophys. J. Suppl.} {\bf 148}, 213 (2003)

\bibitem{weinb03}
S. Weinberg.
\newblock website: www.amacad.org/publications/bulletin/summer2003/weinberg.pdf

\bibitem{bosegr}
S. Bose and L.P. Grishchuk.
\newblock {\em Phys. Rev. D}{\bf 66}, 043529 (2002).

\bibitem{sw}
R.~K. {Sachs} and A.~M. {Wolfe}.
\newblock {\em Astrophys. J.} {\bf 147}, 73 (1967).

\bibitem{dautc}
G. Dautcourt
{\em Month. Not. R. Astr. Soc.} {\bf 144}, 255 (1969)

\bibitem{grzel}
L. P. Grishchuk and Ya. B. Zeldovich.
\newblock {\em Astron. Zh.} {\bf 55}, 209 (1978) [Sov. Astron. {\bf 22}, 125
(1978)].

\bibitem{poln}
A. G. Polnarev.
\newblock Sov. Astron. {\bf 29}, 607 (1985)

\bibitem{Chandrasekhar1960}
S.~Chandrasekhar.
\newblock {\em {Radiative Transfer}}. New York, Dover Publications Inc., 1960

\bibitem {Zaldarriaga1997}
M.~{Zaldarriaga} and U.~{Seljak}.
\newblock {\em Phys. Rev. D} {\bf 55}, 1830 (1997)

\bibitem{Kamionkowski1997}
M.~{Kamionkowski}, A.~{Kosowsky}, and A.~{Stebbins}.
\newblock {\em Phys. Rev. D} {\bf 55}, 7368 (1997)

\bibitem{Basko1980}
M.~M. {Basko} and A.~G. {Polnarev}.
\newblock {\em Sov. Astron.} {\bf 24}, 268 (1980)

\bibitem{Seljak1997a}
U.~{Seljak}.
\newblock {\em Astrophys. J.} {\bf 482}, 6 (1997)

\bibitem{Hu1997b}
W.~{Hu} and M.~{White}.
\newblock {\em New Astron.} {\bf 2}, 323 (1997)

\bibitem{yaglom}
A. M. Yaglom.
\newblock {\em An Introduction to the Theory of Stationary Random Functions}.
Prentice-Hall, Englewood Cliffs, NJ, 1962

\bibitem{grmar}
L.~P. {Grishchuk} and J.~{Martin}.
\newblock {\em Phys. Rev. D}{\bf 56}, 1924 (1997)

\bibitem{cct}
R. Crittenden, D. Coulson, and N. Turok.
\newblock {\em Phys. Rev. D}{\bf 52}, R5402 (1995)

\bibitem{gr96}
L. P. Grishchuk.
\newblock {\em Phys. Rev. D}{\bf 53}, 6784 (1996)

\bibitem{Planck}
{The Planck Collaboration}.
\newblock {\em arXiv:astro-ph/0604069}

\bibitem{bicep1}
B.~{Keating} and N.~{Miller}.
\newblock {\em New Astron. Review,} {\bf 50}, 184 (2006)

\bibitem{bicep2}
J. M. Kovac and D. Barkats
{\em CMB from the South Pole: Past, Prsent, and Future}, 
arXiv:0707.1075

\bibitem{clover}
A. Taylor {\it et al.}
\newblock In: {\it Proceedings of 39th Recontres de Moriond} (Frontieres, 2004)

\bibitem{QUIET}
C.~Lawrence.
{\em The QUIET Experiment}, (http://hea.iki.rssi.ru/Z-90/). Invited talk at 
the conference ``Zeldovich-90", Moscow, 2004

\bibitem{zbg1}
W. Zhao, D. Baskaran, and L. P. Grishchuk. 
{\em Phys. Rev. D}{\bf 79}, 023002 (2009)

\bibitem{zbg2}
W. Zhao, D. Baskaran, and L. P. Grishchuk. 
{\em Phys. Rev. D}{\bf 80}, 083005 (2009)


\end{thebibliography}
%


\end{document}